\documentclass[11pt,a4paper]{article}
\usepackage{geometry}
\usepackage{setspace}

\usepackage{blindtext}
\usepackage[T1]{fontenc}
\usepackage[utf8]{inputenc}
\usepackage{mathptmx}
\usepackage{amssymb}
\usepackage{amsmath}
\usepackage{amsthm}
\usepackage{amsfonts}
\usepackage{chemformula}
\usepackage{cite}
\usepackage[colorlinks, linkcolor=black, anchorcolor=black, citecolor=black]{hyperref}
\usepackage{graphicx}
\usepackage{pdfpages}
\usepackage{pythontex}
\usepackage{import}

\usepackage{caption}
\usepackage{subcaption}
\numberwithin{figure}{section}
\numberwithin{table}{section}

\usepackage{listings}

\usepackage{xspace}

\begin{document}


\includepdf[pages=-]{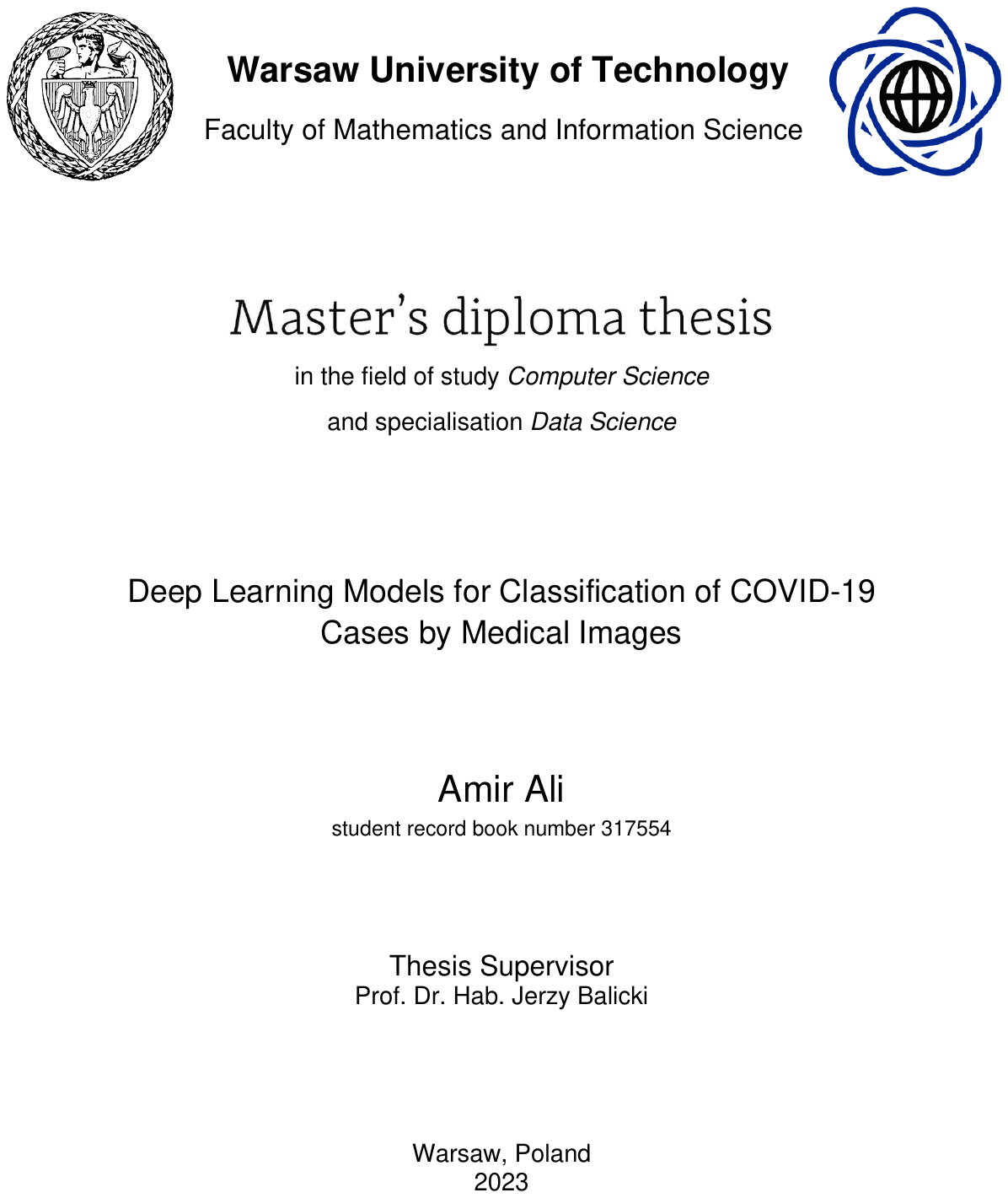}
\pagenumbering{roman}


\begin{center}
\textbf{Deep Learning Models for Classification of COVID-19 Cases by Medical Images}
\end{center}

\textbf{Abstract: }In recent times, the use of chest Computed Tomography (CT) images for detecting coronavirus infections has gained significant attention, owing to their ability to reveal bilateral changes in affected individuals. However, classifying patients from medical images presents a formidable challenge, particularly in identifying such bilateral changes. To tackle this challenge, our study harnesses the power of deep learning models for the precise classification of infected patients. Our research involves a comparative analysis of deep transfer learning-based classification models, including DenseNet201, GoogleNet, and AlexNet, against carefully chosen supervised learning models. Additionally, our work encompasses Covid-19 classification, which involves the identification and differentiation of medical images, such as X-rays and electrocardiograms, that exhibit telltale signs of Covid-19 infection. This comprehensive approach ensures that our models can handle a wide range of medical image types and effectively identify characteristic patterns indicative of Covid-19. By conducting meticulous research and employing advanced deep learning techniques, we have made significant strides in enhancing the accuracy and speed of Covid-19 diagnosis. Our results demonstrate the effectiveness of these models and their potential to make substantial contributions to the global effort to combat COVID-19. \newline

\clearpage

\begin{center}
\textbf{Modele uczenia głębokiego do klasyfikacji przypadków Covid-19 za pomocą obrazów medycznych}
\end{center}

\textbf{Streszczenie: }Wykorzystanie obrazów tomografii komputerowej klatki piersiowej (CT) do wykrywania zakażeń koronawirusem jest istotne, ze względu na efektywno ujawniania zmian obustronnych u chorych osób. Klasyfikacja pacjentów na podstawie obrazów medycznych stanowi ogromne wyzwanie, zwłaszcza w identyfikacji zmian obustronnych. Aby zmierzyć się z tym wyzwaniem, zastosowano model głębokiego uczenia maszynowego do klasyfikacji pacjentów. Badania obejmują analizę porównawczą modeli klasyfikacji opartych na transferze wiedzy, takich jak DenseNet201, GoogleNet i AlexNet. Klasyfikację COVID-19, polega na identyfikacji i różnicowaniu obrazów medycznych, takich jak obrazy rentgenowskie i elektrokardiogram. To kompleksowe podejście sprawia, że modele mogą klasyfikowa szeroki zakres rodzajów obrazów medycznych i efektywnie identyfikować charakterystyczne wzorce wskazujące na obecność COVID-19. Dzięki wykorzystaniu zaawansowanych technik głębokiego uczenia się, znaczco poprawiono dokładnoś i szybkoś diagnozowania COVID-19. Uzyskane wyniki dowodzą skuteczności modeli i ich potencjału odnośnie przeciwdziałania COVID-19.\newline
\textbf{Słowa kluczowe:} Diagnostyka Covid-19, Uczenie Maszynowe, Klasyfikacja
\clearpage



**Student Name and Surname**

*Amir Ali*

**Number of Album**

*317554*

**Field of Study**

*Computer Science with specialisation in Data Science*

## DECLARATION

Aware of the criminal responsibility for making false statements, I declare that this thesis has been written by me independently, under the supervision of Professor Dr. Jerzy Balicki.

At the same time, I declare that:

- This thesis does not infringe copyright as defined by the Act of 4 February 1994 on Copyright and Related Rights (Journal of Laws 2006. No. 90, item 631 as amended) and personal rights protected by civil law,

- This thesis does not contain data or information that I have obtained by unauthorized means,

- This thesis has not previously been the basis of any other official procedure for the award of a diploma or professional title,

- All information included in this thesis, obtained from written and electronic sources, has been documented in the literature list with appropriate references,

- I am familiar with the regulations of the Warsaw University of Technology on the management of copyright and related rights, industrial property rights, and the principles of related rights, industrial property rights, and the principles of commercialization.

I declare that the content of the diploma paper in a printed version, the content of the diploma paper contained on an electronic carrier (compact disc) and the content of the diploma thesis in the APD module of the USOS system are identical.

*AMIR*

................................

Student Signature

miejscowosc i data
Warszawa, 19 pazdziernika 2023 r.

**imie i nazwisko studentae**
*Amir Ali*

**numer albumu**
*317554*

**kierunek studiow**
*Informatyka ze specjalizacją Data Science*

## OŚWIADCZENIE

Świadomy odpowiedzialności karnej za składanie fałszywych zeznań oświadczam, że niniejsza praca dyplomowa została napisana przeze mnie, pod kierunkiem prof. dr hab. Jerzy Balicki
Jednocześnie oświadczam, że:

- niniejsza praca dyplomowa nie narusza praw autorskich w rozumieniu ustawy z dnia 4 lutego 1994 roku o prawie autorskim i prawach pokrewnych (Dz.U. z 2006 r. Nr 90, poz. 631 z późn. zm.) oraz dóbr osobistych chronionych prawem cywilnym,

- niniejsza praca dyplomowa nie zawiera danych i informacji, które uzyskałem/-am w sposób niedozwolony,

- niniejsza praca dyplomowa nie była wcześniej podstawą żadnej innej urzędowej procedury związanej z nadawaniem dyplomów lub tytułów zawodowych

- wszystkie informacje umieszczone w niniejszej pracy, uzyskane ze źródeł pisanych i elektronicznych, zostały udokumentowane w wykazie literatury odpowiednimi odnośnikami,

- znam regulacje prawne Politechniki Warszawskiej w sprawie zarządzania prawami autorskimi i prawami pokrewnymi, prawami własności przemysłowej oraz zasadami komercjalizacji.

Oświadczam, że treść pracy dyplomowej w wersji drukowanej, treść pracy dyplomowej zawartej na nośniku elektronicznym (płycie kompaktowej) oraz treść pracy dyplomowej w module APD systemu USOS są identyczne.

*AMIR*
........................................
czytelny podpis studenta




\renewcommand{\contentsname}{Table of Contents}
\setcounter{tocdepth}{3}
\tableofcontents
\addtocontents{toc}{\par\nobreak \mbox{}\hfill{\bf Page}\par\nobreak}
\clearpage


\pagenumbering{arabic}

\section{Introduction}\label{sec:intro}
In this chapter, we'll address the following key components: defining the problem statement, sharing personal motivation, outlining research aims and objectives, and in the end summarizing the content.
\subsection{Problem Statement}
The COVID-19 pandemic has led to an unparalleled global public health crisis, impacting millions of individuals on a worldwide scale. In addition to the direct impact on human health, the pandemic has also led to significant social and economic disruptions. One of the key challenges in managing the pandemic is the accurate and timely diagnosis of COVID-19 cases\cite{shi2020overview}.

Medical imaging has risen as a promising method for diagnosing COVID-19. Commonly, chest X-rays and CT scans are employed to detect the distinctive attributes of COVID-19 pneumonia, like ground-glass opacities and consolidation. Nevertheless, the analysis of medical images is subjective and can differ substantially among radiologists. This divergence can result in diagnostic inaccuracies and treatment setbacks, carrying potential grave repercussions for patients.

Deep learning models offer a potential solution to the challenges of medical image interpretation. These models can learn to identify patterns in medical images that are indicative of COVID-19 pneumonia, without the need for human interpretation. However, existing deep learning models for COVID-19 diagnosis have limitations such as low accuracy and limited generalizability. More research is needed to develop accurate and reliable models that can be used in clinical practice\cite{shorten2021deep}.

Hence, the primary objective of this thesis is to create advanced deep-learning models tailored to categorize instances of COVID-19 through the utilization of medical images. The specific goals of this study encompass amassing a comprehensive dataset of medical images, refining the images through preprocessing techniques, training and assessing the effectiveness of deep learning models, and subsequently contrasting the achievements of diverse models.

The outcomes of this investigation carry substantial significance for the identification and supervision of COVID-19 cases. Furthermore, these findings have the potential to contribute significantly to the evolution of enhanced and streamlined diagnostic instruments, promising more efficient and accurate diagnostic procedures.

\subsection{Personal Motivation}
My longstanding fascination has revolved around the utilization of deep learning within the healthcare sector. More specifically, I've been captivated by the inherent potential of deep learning models to enhance both the precision and swiftness of scrutinizing medical images. Given that medical imaging occupies a pivotal role in the identification and therapeutic interventions of numerous illnesses, the capability to automate the assessment of medical images bears the potential to wield a considerable influence on the overall outcomes of healthcare practices.

The COVID-19 pandemic has brought these issues into sharp focus, as the rapid and accurate diagnosis of COVID-19 cases is essential for effective management and control of the disease. Medical imaging, such as chest X-rays and CT scans, has emerged as a valuable tool for the diagnosis of COVID-19, but the interpretation of medical images can be challenging. Deep learning models offer a promising solution to these challenges, as they can learn to identify patterns in medical images that are indicative of COVID-19 pneumonia.

The potential of deep learning models for COVID-19 diagnosis has been demonstrated in several recent studies, but there is still much work to be done to develop accurate and reliable models that can be used in clinical practice. As a student of master's in data science, I am excited about the opportunity to contribute to this field by developing and evaluating deep-learning models for the classification of COVID-19 cases using medical images.

Moreover, the opportunity to work on a project that has the potential to make a positive contribution to the fight against the COVID-19 pandemic is deeply motivating to me. The pandemic has had a profound impact on society, and the ability to develop more accurate and efficient diagnostic tools could have a significant impact on healthcare outcomes and public health response. I am committed to using my skills and knowledge to contribute to this effort in any way that I can.

Working on this project will also provide me with valuable experience in conducting research, collecting and analyzing data, and developing deep learning models. These skills will be transferable to future research projects and could also be useful in my future career. I am highly motivated to undertake this research and to make a positive contribution to the field of deep learning for medical imaging and COVID-19 diagnosis.

\subsection{Research Aims and Objectives}
The aim of developing the classification and detection system is to create a reliable and accurate method with a tool for identifying COVID-19 cases from medical images. This system can assist in diagnosing and treating patients more efficiently and effectively, particularly in areas where access to specialized medical professionals or testing equipment is limited. To achieve this aim, the following objectives will be pursued:

\begin{enumerate}
    \item To conduct a comprehensive review of the literature on deep learning models for medical image analysis and COVID-19 diagnosis, with a particular focus on models that have been developed for the classification of COVID-19 cases using medical images.

    \item To collect and curate a dataset of medical images for the classification of COVID-19 cases. This dataset will include chest X-rays and CT scans from a variety of sources, and will be annotated with labels indicating the presence or absence of COVID-19.

    \item To develop and train deep learning models for the classification of COVID-19 cases using medical images. The models will be based on state-of-the-art deep learning architectures, and will be trained using the curated dataset.

    \item To evaluate the performance of the developed models on the classification of COVID-19 cases using medical images. The models will be evaluated using standard metrics for classification, such as accuracy, precision, loss, and f1-score.

    \item To compare the performance of the developed models with existing methods for COVID-19 diagnosis using medical images, such as radiologist interpretation and other deep learning models.

    \item To conduct a sensitivity analysis to investigate the effect of different parameters and hyperparameters on the performance of the developed models.
    \end{enumerate}
    
The primary aim of this study is to make a meaningful contribution towards the advancement of diagnostic tools for COVID-19 that are not only more precise but also highly efficient. Such progress holds the potential to exert a substantial influence on healthcare outcomes and the broader public health strategy. Furthermore, this research endeavor will furnish valuable insights into the efficacy of employing deep learning models in the analysis of medical images. These insights could prove instrumental in shaping the creation of analogous models for different diseases and various applications beyond the scope of COVID-19.

\subsection{Summary}
The thesis structure for a research project on Deep learning models for the classification of Covid-19
cases by medical images include the following sections:

\begin{enumerate}

    \item \textbf{Introduction:} This section provides an overview of the problem statement, personal motivation, research aim and objectives for the thesis.

    \item \textbf{Literature Review:} This section provides a succinct overview of existing deep learning models in medical image analysis and their relevance to COVID-19 diagnosis. The discussion includes the limitations of current methods and a brief review of COVID-19.

    \item \textbf{System Architecture:} This combined sections describes the data collection process and dataset used for the study, along with the data preprocessing steps. It then outlines the proposed methodology, including the selection of deep learning architectures, training procedures, and evaluation strategies for COVID-19 diagnosis using medical images.

    \item \textbf{Result Evaluation:} This section presents the outcomes of the developed models, analyzing their performance using standard classification metrics and comparing them with existing methods. It showcases the effectiveness of the deep learning models for accurate COVID-19 diagnosis from medical images.

     \item \textbf{Conclusion and Future Work:}  Combining these sections, the conclusion summarizes the study's key findings, discussing the potential of deep learning models for COVID-19 diagnosis using medical images. It also outlines future research directions, including the exploration of transfer learning and extending the models to diagnose other diseases. Challenges and limitations are addressed, providing insights for further research in medical image analysis.

\end{enumerate}

\subsection{Remarks and Conclusions}
In the midst of the global COVID-19 pandemic, the development of precise and efficient diagnostic tools has become paramount. Our research, which focuses on the application of deep learning to analyze medical images, represents a significant step towards meeting this need.

The pandemic has emphasized the importance of swift and accurate diagnosis, with medical imaging emerging as a valuable tool. However, the subjectivity and variability in image interpretation by human experts have underscored the necessity for automated and objective methods. Deep learning models offer a promising solution by learning to identify patterns in medical images indicative of COVID-19 pneumonia.

Our study aimed to address the limitations of existing models and develop accurate, reliable, and efficient deep learning models for COVID-19 diagnosis using medical images. Through the collection and curation of a comprehensive dataset, extensive preprocessing, model development, and rigorous evaluation, we have made significant progress in this direction.
\clearpage
\section{Literature Review}\label{sec:lit-rev}
In this chapter, we will address the following topics: COVID-19 Origin and Nature, Radiographic Diagnosis for COVID-19, Deep Learning for COVID-19 Classification, Deep Learning Applications in Diagnosis and Treatment, and finally, provide remarks and conclusions.
\subsection{COVID-19 Origin and Nature}
\subsubsection{Overview of COVID-19 }
The SARS-CoV-2 virus, initially identified in December 2019 in Wuhan, China, has led to the continuous global health crisis known as the COVID-19 pandemic. It has since spread rapidly across the world, with the World Health Organization (WHO) declaring it a pandemic in March 2020\cite{ciotti2020covid}. Figure 2.1 shows the structure of the SARS-CoV-2 virus, responsible for COVID-19, showing its genetic material enclosed in a lipid membrane with spike proteins on the surface. 

\begin{figure}[ht]
    \centering
    \includegraphics[width=0.4\textwidth]{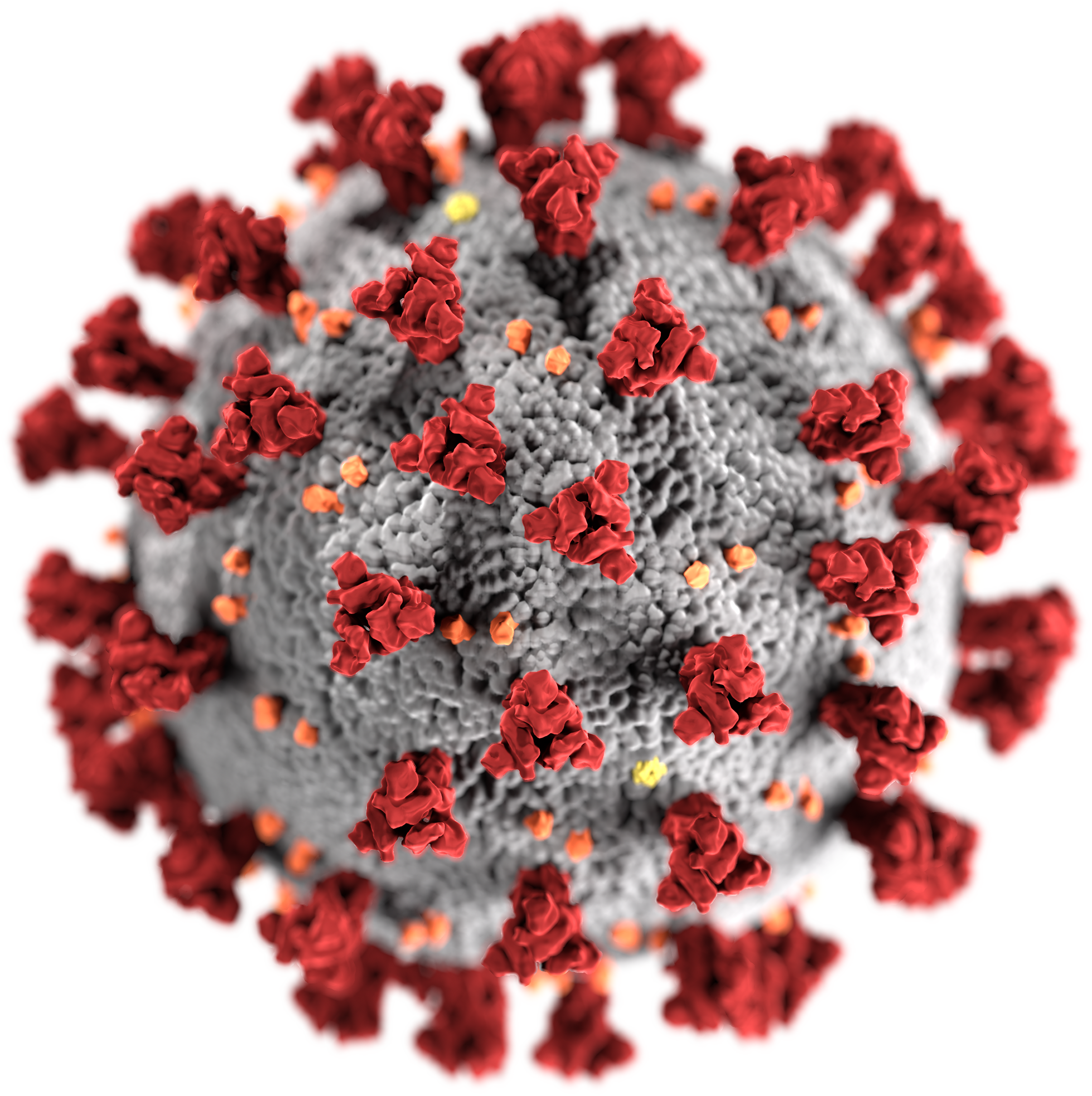}
    \caption{SARS-CoV-2\cite{wikipediaCOVID19Wikipedia}}
\end{figure}

The transmission of the virus mainly occurs through respiratory droplets released when an individual who is infected talks, coughs, or sneezes. Additionally, transmission can also happen by touching surfaces that have been contaminated with the virus and subsequently touching the face. The symptoms associated with COVID-19 can vary in intensity, encompassing mild to severe manifestations such as fever, cough, fatigue, and difficulty breathing. In certain instances, the virus has the potential to induce serious respiratory ailments, pneumonia, and even fatalities.\cite{aabed2021analytical}.

As of September 2021, global COVID-19 cases have surpassed 229 million, resulting in a staggering death toll exceeding 4.7 million. The pandemic has had far-reaching social, economic, and political impacts, including widespread lockdowns, travel restrictions, and disruptions to daily life.

In response to the pandemic, numerous measures have been implemented, including the development and distribution of vaccines, increased testing and contact tracing, and the use of personal protective equipment (PPE) to prevent transmission\cite{guner2020covid}.

The COVID-19 pandemic has spurred significant research efforts into the virus, its transmission, and its impact on human health. In particular, there has been a focus on developing effective treatments and vaccines to mitigate the effects of the pandemic and prevent future outbreaks. Additionally, the pandemic has highlighted the importance of public health preparedness and the need for coordinated global responses to address emerging infectious diseases\cite{harper2020impact}.

\subsubsection{Origin of COVID-19}

The origin of COVID-19, the disease caused by the SARS-CoV-2 virus, is still being studied and debated among the scientific community. However, it is widely believed that the virus originated from bats and possibly passed through an intermediate host animal before infecting humans\cite{umakanthan2020origin}.

In December 2019, the earliest instances of COVID-19 were documented in Wuhan, China. Initially, the hypothesis emerged that the virus might have emerged from a local seafood market where live animals were traded. Nonetheless, subsequent investigations indicate that the market potentially functioned as an amplification point for the virus rather than serving as its point of origin.

Genetic analysis of the virus has shown that it is closely related to other coronaviruses found in bats, particularly those found in horseshoe bats in China. The prevailing notion suggests that the virus could have transitioned from bats to an intermediary host, possibly a pangolin, before ultimately transmitting to humans. Figure 2.2 shows how COVID-19 is transmitted from the primary host to humans.

\begin{figure}[ht]
    \centering
    \includegraphics[width=1\textwidth]{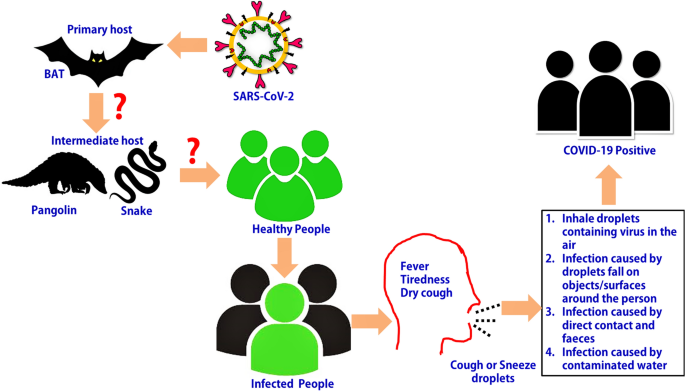}
    \caption{SARS-CoV-2\cite{mishra2020overview}}
\end{figure}

Despite numerous investigations into the origin of the virus, there is still much uncertainty and speculation surrounding the exact source and route of transmission. In May 2021, the World Health Organization released a report on its investigation into the origins of the virus, concluding that it was "extremely unlikely" that the virus was accidentally released from a laboratory and that it was more likely to have originated from the animal-to-human transmission. However, some experts continue to call for further investigation and research into the origins of the virus\cite{bloom2021investigate}.

\subsubsection{Structure of COVID-19}
The virus is spherical in shape and has a diameter of approximately 120 nanometers. The virus consists of different components: Spike, Nucleocapsid, Membrane, Envelope, and RNA viral genome \cite{schoenmaker2021mrna}. Figure 2.3 shows the structure of COVID-19.

\begin{figure}[ht]
    \centering
    \includegraphics[width=0.7\textwidth]{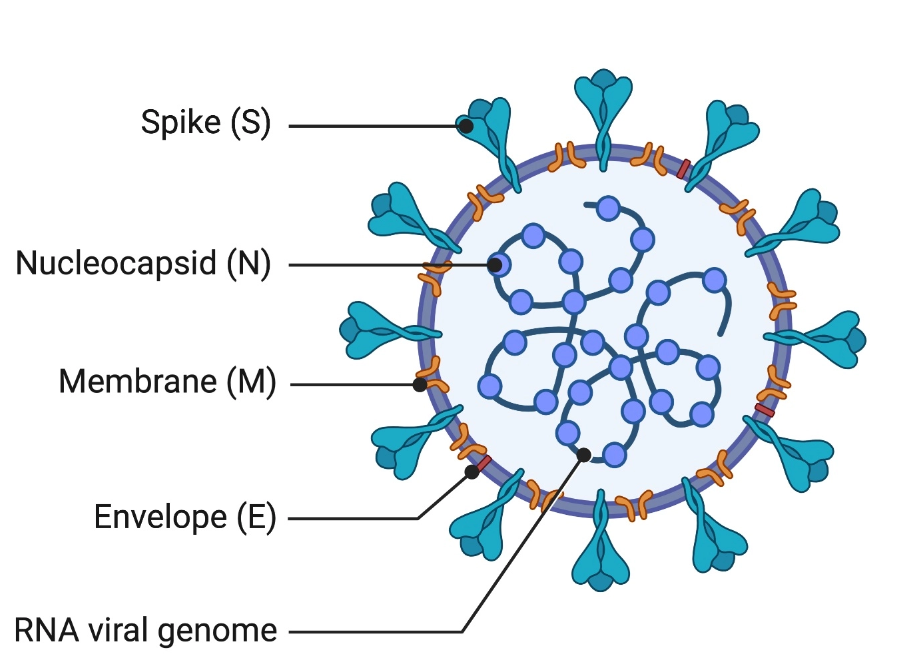}
    \caption{Structure of Covid-19\cite{king2020coronavirus}}
\end{figure}

The spike protein is located on the outer surface of the virus and plays a crucial role in the virus's ability to infect host cells. The spike protein allows the virus to attach to and enter host cells by binding to specific receptors on the surface of the cell. This interaction between the spike protein and host cell receptors is the initial step in the virus's entry into the cell and is a key target for antiviral therapies.

The nucleocapsid is the protein shell that encapsulates the viral RNA genome. It plays a critical role in protecting the viral genome from degradation and facilitating its replication and transcription. The nucleocapsid also interacts with host cell proteins to facilitate virus replication and can trigger an immune response in the host.

The viral envelope is a lipid bilayer that surrounds the nucleocapsid and contains various viral proteins, including the spike protein. The envelope is derived from the host cell membrane as the virus buds out of the infected cell. It is important for protecting the virus from the host immune system and aiding in the virus's entry into host cells.

The viral membrane is a matrix protein that lies between the nucleocapsid and the envelope. It helps to maintain the structural integrity of the virus and is involved in viral assembly and release.

The RNA viral genome is the genetic material of the virus and is composed of a single strand of RNA. The viral genome contains all the genetic information necessary for the virus to replicate and produce its proteins. The RNA genome of the COVID-19 virus is approximately 30,000 nucleotides long and encodes for 29 proteins. \newpage

Together, these components make up the structure of the COVID-19 virus and play critical roles in the virus's ability to infect host cells, replicate, and cause disease. Understanding the structure and function of these components is essential for the development of effective treatments and vaccines to combat COVID-19.

\subsubsection{Immune Responses To COVID-19}

COVID-19 elicits a complex immune response in the human body, involving various components of the immune system such as T cells, B cells, and antibodies. Upon infection, the virus activates the innate immune response, which is the first line of defense against pathogens. The innate immune response is initiated by the detection of viral components by the host cells, which trigger the production of interferons and pro-inflammatory cytokines. These molecules recruit immune cells to the site of infection and induce an antiviral state in neighboring cells \cite{chowdhury2020immune}.

As the infection progresses, the adaptive immune response is activated, which is more specific and targeted than the innate response. The adaptive response involves the activation of T and B cells, which can recognize and target specific components of the virus. CD4+ T cells, also known as helper T cells, play a critical role in coordinating the immune response by activating other immune cells and promoting antibody production. CD8+ T cells, also known as cytotoxic T cells, directly target and kill infected cells \cite{scully2020considering}.

B cells, on the other hand, produce antibodies that can neutralize the virus and prevent its entry into host cells. Antibodies can also tag the virus for destruction by other immune cells. The spike protein of the virus, which is responsible for binding to the host cell receptor, is the primary target of neutralizing antibodies \cite{sewell2020cellular}.

However, the immune response to COVID-19 can also lead to excessive inflammation and tissue damage, especially in severe cases. This phenomenon, known as a cytokine storm, is characterized by the overproduction of pro-inflammatory cytokines and can lead to multiple organ failures and even death. The cytokine storm is thought to be caused by an uncontrolled immune response, which can be exacerbated by pre-existing conditions such as diabetes, hypertension, and obesity \cite{shi2020covid}.

Understanding the immune response to COVID-19 is critical for the development of effective vaccines and therapeutics. Vaccines can elicit a protective immune response by priming the immune system to recognize and respond to the virus, while therapeutics can modulate the immune response to prevent excessive inflammation and tissue damage.

\subsubsection{Variants of COVID-19}

As the COVID-19 virus continues to spread globally, it has undergone mutations or changes in its genetic makeup. These changes result in the emergence of new variants or strains of the virus. Variants of COVID-19 are named after the place of their origin or the mutations they carry. Some of the most well-known variants of COVID-19 include the Alpha, Beta, Gamma, and Delta variants \cite{duong2021s}.

The Alpha variant (B.1.1.7), which was initially detected in the United Kingdom in September 2020, is believed to possess a transmissibility that surpasses that of the original COVID-19 virus by up to 70 percent. This variant displays numerous mutations in the virus's spike protein, enabling it to exhibit enhanced affinity for human cells and facilitating more efficient infection. The Alpha variant has spread rapidly and has been reported in several countries worldwide \cite{eyre2022effect}. \newpage

The Beta variant (B.1.351), first identified in South Africa, has mutations in the spike protein of the virus that help it to evade the immune system's response. This variant is thought to be less sensitive to some of the COVID-19 vaccines currently available. The Beta variant has also been reported in several countries worldwide.

The Gamma variant (P.1), first identified in Brazil, has mutations in the spike protein of the virus that may make it more transmissible and less susceptible to some of the antibodies produced by the immune system. The Gamma variant has been reported in several countries worldwide and is associated with a surge in cases in Brazil \cite{duong2021alpha}.

The Delta variant (B.1.617.2), first identified in India, has multiple mutations in the spike protein of the virus that make it more transmissible than the original virus. The Delta variant is now the dominant strain of COVID-19 worldwide and has been reported in many countries, including the United States, the United Kingdom, and Australia. The Delta variant has been associated with a surge in cases and hospitalizations in many countries and is believed to be more severe than previous variants \cite{yadav2021neutralization}.

Other variants of COVID-19 have also been identified, including the Epsilon, Zeta, Eta, Theta, and Kappa variants. These variants have different mutations in the spike protein of the virus, which may affect their transmissibility, severity, or response to vaccines \cite{abdool2021appropriate}.

It is essential to continue monitoring the emergence of new variants of COVID-19 and understanding their properties to inform public health strategies and vaccine development \cite{rubin2021covid}\cite{darby2021covid}.

\subsection{Radiographic Diagnosis for COVID-19}
Radiographic diagnostic methods, such as chest X-ray, Computed Tomography (CT) scan, Ultrasound, and Magnetic Resonance Imaging (MRI) are used for the diagnosis and monitoring of COVID-19. These methods help to identify the characteristic lung abnormalities associated with the disease \cite{kaufman2020review}.
\subsubsection{Chest X-ray for COVID-19}
Chest X-ray is one of the widely used radiographic diagnostic methods for COVID-19 diagnosis. The disease causes pneumonia and affects the lungs, making chest X-rays an essential tool in its diagnosis. The X-ray image shows the lungs' condition, and radiologists can identify the presence of inflammation, fluid buildup, or any other abnormalities in the lungs caused by COVID-19. Figure 2.4 displays an X-ray image showcasing COVID-19-related findings in a patient's chest.

\begin{figure}[ht]
    \centering
    \includegraphics[width=0.35\textwidth]{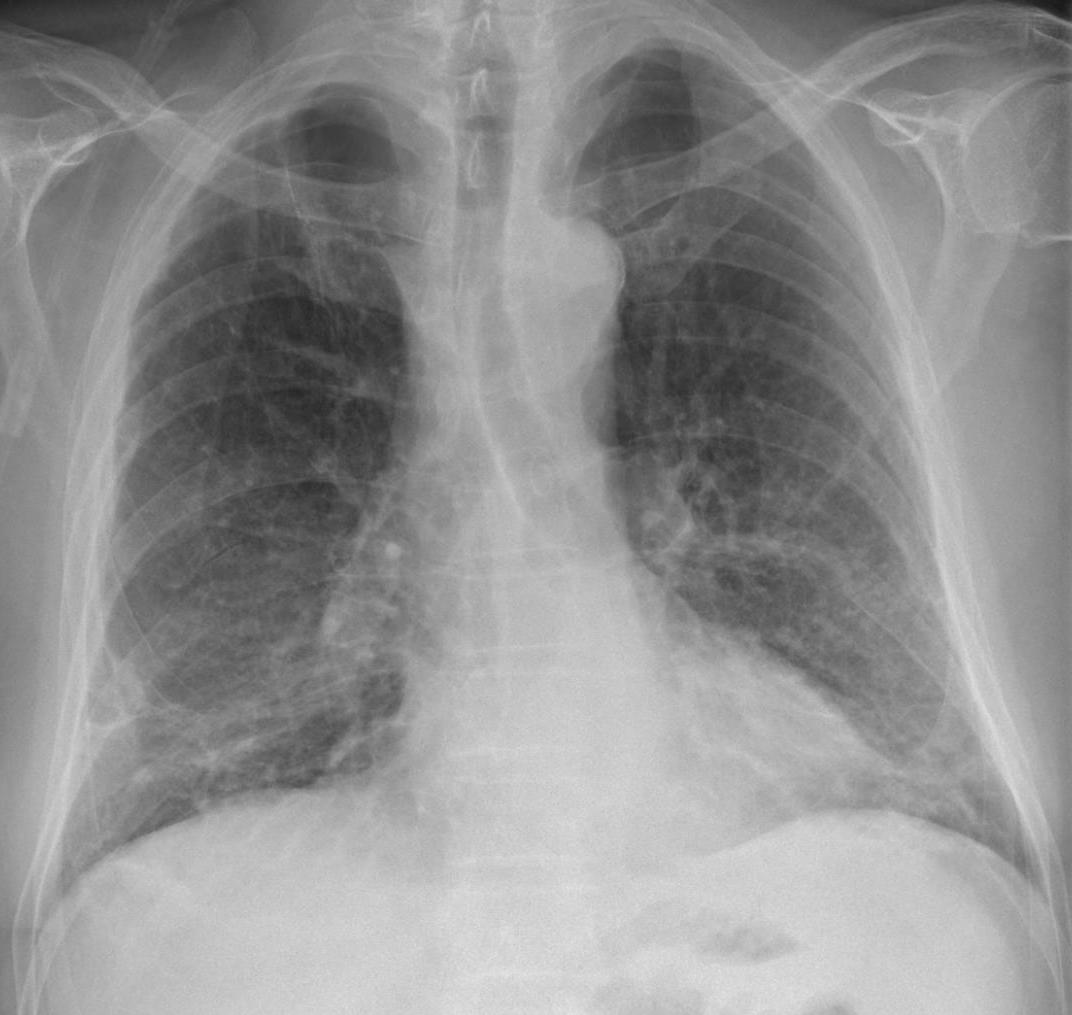}
    \caption{X-ray of Covid-19 Patient\cite{mobihealthnewsBoltonPartners}}
\end{figure}

\newpage

The COVID-19 pneumonia pattern on a chest X-ray appears as ground-glass opacities (GGOs), consolidations, or a combination of both. The radiological findings are similar to those observed in other types of viral pneumonia, making it challenging to differentiate between them. However, the radiographic features of COVID-19 pneumonia are distinct, and radiologists can make an accurate diagnosis with the help of a chest X-ray \cite{mangal2020covidaid}. 

\subsubsection{CT Scans for COVID-19}

Computed tomography (CT) scan is a radiographic diagnostic method that uses X-rays and computer processing to create detailed images of the body's internal structures. CT scan is a valuable tool for diagnosing COVID-19 as it can detect lung abnormalities, such as ground-glass opacities and consolidation, which are characteristic of COVID-19 pneumonia \cite{zhao2020covid}.

CT scans can provide important diagnostic information in patients with COVID-19, especially those with a severe or critical illness. It can also be useful in identifying complications of COVID-19, such as pulmonary embolism, pleural effusion, and pneumothorax. Figure 2.5 presents CT scans illustrating COVID-19-related findings in a patient.

\begin{figure}[ht]
    \centering
    \includegraphics[width=0.4\textwidth]{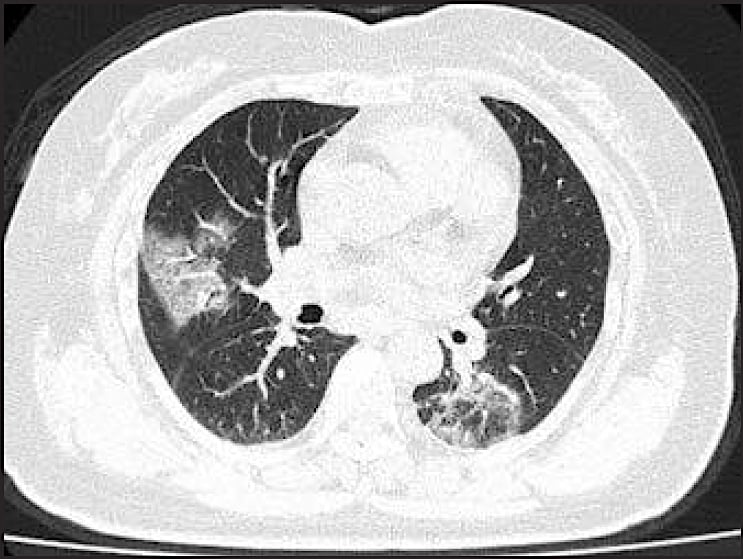}
    \caption{CT scans of Covid-19 Patient\cite{itnonlineRadiologyLessons}}
\end{figure}

However, CT scan has some limitations and risks, such as radiation exposure and the potential for false negatives in patients with early or mild disease. Therefore, CT scans should be used judiciously in the diagnosis and management of COVID-19, in combination with other clinical and laboratory findings.

\subsubsection{Ultrasound for COVID-19}

Ultrasound has also emerged as a potential diagnostic tool for COVID-19, particularly in cases where patients may not be able to undergo a CT scan or X-ray. Ultrasound machines are portable, affordable, and non-invasive, making them an attractive option for point-of-care diagnosis in areas with limited resources or for patients who cannot tolerate other imaging modalities. Figure 2.6 showcases an ultrasound image associated with a COVID-19 patient, offering insights into disease-related observations or conditions. \newpage

\begin{figure}[ht]
    \centering
    \includegraphics[width=0.35\textwidth]{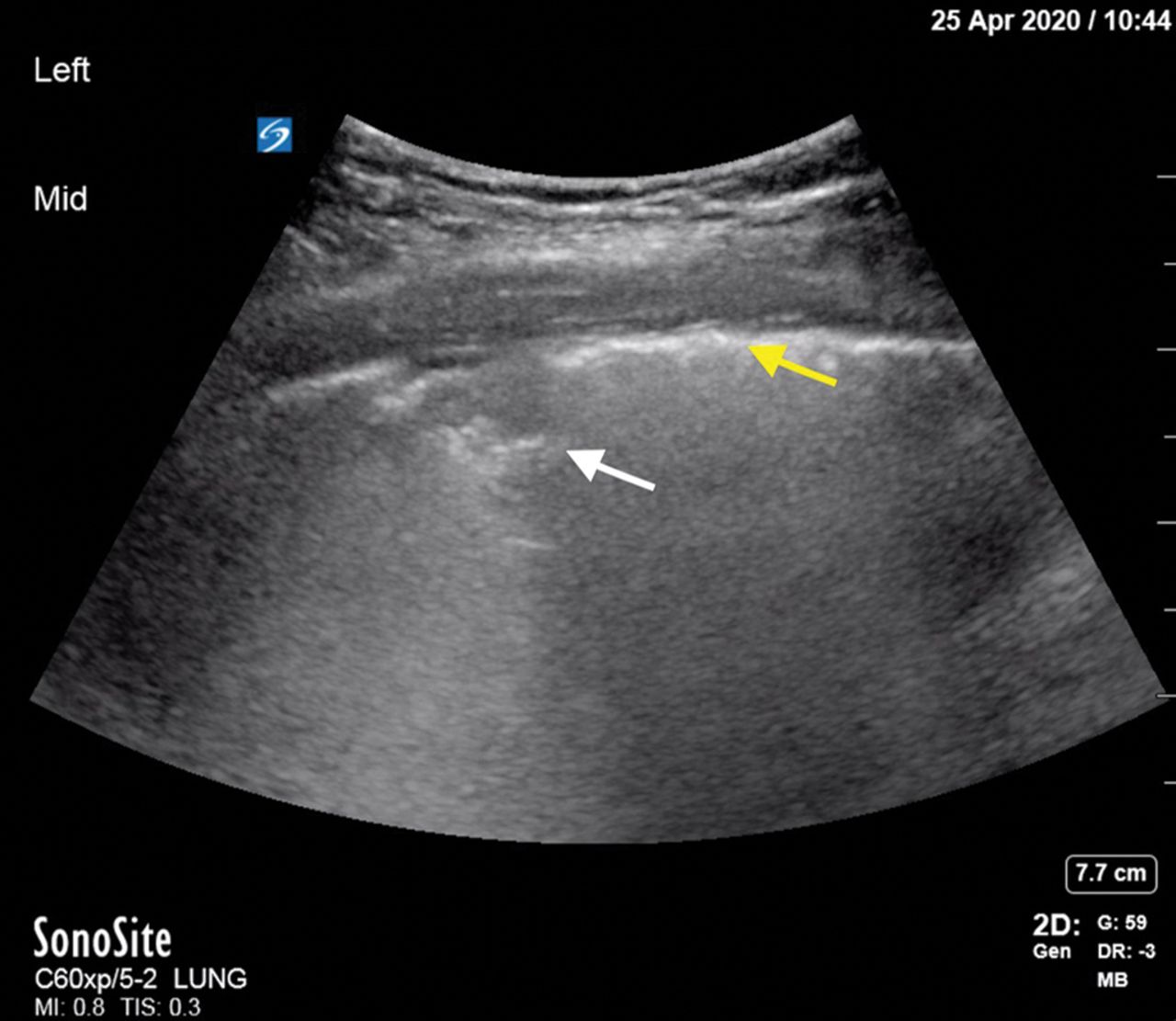}
    \caption{Ultrasound of Covid-19 Patient\cite{jackson2021lung}}
\end{figure}
The use of ultrasound for COVID-19 diagnosis typically involves the examination of the lungs, where the virus can cause characteristic findings such as thickened pleural lines, consolidations, and B-lines. However, ultrasound has lower sensitivity and specificity compared to CT scans and chest X-rays for COVID-19 diagnosis, and its use is still being evaluated in clinical studies \cite{convissar2019application}.

\subsubsection{MRI for COVID-19}

Magnetic resonance imaging (MRI) serves as a non-intrusive diagnostic instrument employing a powerful magnetic field and radio waves to generate comprehensive visuals of the body's internal structures. It is increasingly being explored as a potential diagnostic tool for COVID-19, particularly in cases where CT scans and X-rays do not provide conclusive results. MRI is particularly useful in detecting lung abnormalities such as pneumonia and pulmonary fibrosis, which are commonly associated with severe COVID-19 infections \cite{bispo2022brain}.

One of the main advantages of MRI is that it does not expose patients to ionizing radiation, which is a potential risk associated with CT scans and X-rays. This makes MRI a safer option for patients who require multiple imaging tests or who are at increased risk of developing radiation-induced cancers. However, MRI scans can take longer to perform and may not be as widely available as other imaging tests. Additionally, MRI machines are expensive and require specialized training to operate, which may limit their use in certain healthcare settings. Figure 2.7 displays an MRI scan of a COVID-19 patient, potentially revealing specific medical insights related to the disease.

\begin{figure}[ht]
    \centering
    \includegraphics[width=0.4\textwidth]{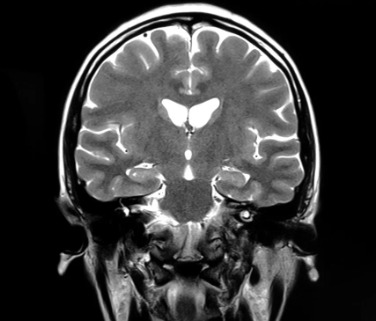}
    \caption{MRI of Covid-19 Patient\cite{newsmedicalShowsBrain}}
\end{figure}

\subsection{Deep Learning for COVID-19 Classification}
There are several deep-learning models that have been developed to classify COVID-19 images. Some examples include the Convolutional Neural Network (CNN), Residual Neural Network (ResNet), Inception V3, and DenseNet. Each of these models utilizes different architectures and techniques to analyze the images and classify them as either COVID-19 positive or negative \cite{sevi2020covid}.

\subsubsection{Convolutional Neural Network}
Convolutional Neural Networks (CNNs) are a type of deep learning neural network designed for processing and analyzing images. They consist of multiple layers, including convolutional layers, pooling layers, and fully connected layers. CNNs are designed to automatically learn features from images by passing them through several layers of filters, which allows them to recognize patterns and structures in the images \cite{o2015introduction}.

The first layer in a CNN is called a convolutional layer. It uses filters to scan the input image and make maps of different features like edges or corners. These filters are learned while the network trains, getting better at recognizing important details.

The next layer in a CNN is usually a pooling layer. This layer reduces the spatial dimension of the feature maps produced by the convolutional layer by down-sampling the output. This reduces the computational complexity of the network and helps to prevent overfitting.

Following multiple convolutional and pooling layers, the feature maps are flattened out and sent into a fully connected layer. This layer functions as a classifier, utilizing the identified features to make predictions regarding the input image. Figure 2.8 depicts the structure and design of a Convolutional Neural Network, commonly used for image-related tasks in deep learning

\begin{figure}[ht]
    \centering
    \includegraphics[width=1\textwidth]{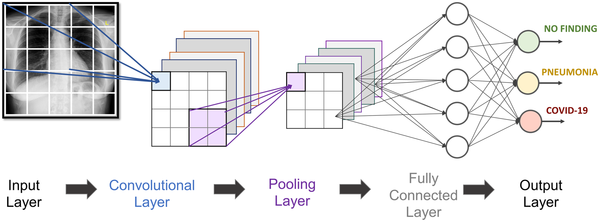}
    \caption{Architecture of CNN\cite{maior2021convolutional}}
\end{figure}

Deep Convolutional Neural Networks (DCNNs) are a type of CNN with multiple layers, typically more than 10 or 20 layers. The deeper architecture allows the network to learn more complex features and patterns from the input data. However, training a DCNN can be challenging due to the vanishing gradient problem, where the gradients become too small to update the weights effectively.
\newpage
To address this issue, DCNNs use techniques such as batch normalization, residual connections, and skip connections. Batch normalization normalizes the input to each layer to reduce the internal covariate shift and help with gradient propagation. Residual connections allow the network to learn residual functions to improve the convergence and gradient flow during training. Skip connections also improve gradient flow by allowing information to bypass some layers and pass directly to deeper layers.

DCNNs have shown remarkable success in various computer vision tasks, such as image recognition, object detection, segmentation, and many others \cite{tulbure2022review}.

\subsubsection{Residual Neural Network}
Residual Neural Networks, commonly referred to as ResNets, were introduced in 2016 as a solution for Image Recognition \cite{he2016deep}. The creators of ResNets put forth a novel method to tackle the issue of vanishing gradients, a challenge that can emerge in extremely deep neural networks.

The vanishing gradient problem arises when the gradient signal that is backpropagated through the network becomes too small to update the weights of the earlier layers effectively. This is particularly problematic in very deep networks, where the gradient signal may need to pass through many layers before reaching the beginning of the network.

ResNets incorporate skip connections to establish shortcuts between layers, which aids in facilitating the smooth flow of gradient signals throughout the network. To be more precise, ResNets introduce a "residual" connection that directly bypasses one or more layers. This connection involves adding the output of an earlier layer to the output of a subsequent layer. This way, if the gradient signal becomes too small to update the weights of the earlier layers effectively, the residual connection can still provide a strong signal for the update.

More formally, a residual connection adds the input \emph{x} to the output \emph{f(x)} of a sequence of layers. Mathematically, this can be written as:

\begin{equation}
y = f(x) + x
\end{equation}

where \emph{y} is the output of the residual block, \emph{f(x)} is the output of the sequence of layers, and \emph{x} is the input to the block. By adding the input to the output, the residual connection effectively learns a residual function that can be added to the input to obtain the desired output. This way, the network can learn to make small changes to the input rather than trying to learn the entire function from scratch.

The architecture of a ResNet typically consists of a series of "residual blocks," each of which contains multiple convolutional layers, followed by batch normalization and ReLU activation functions. The skip connections are added between these residual blocks, allowing the gradient signal to bypass multiple layers and propagate more easily through the network.

ResNets can be very deep, with some architectures containing hundreds of layers. The original ResNet paper reported that a 152-layer ResNet was able to achieve state-of-the-art performance on the ImageNet dataset, which contains over a million images from a thousand different classes. \newpage

Since their inception, ResNets have found extensive application in a variety of computer vision tasks, including image classification, object detection, and segmentation. Notably, ResNets have consistently demonstrated cutting-edge performance across multiple benchmarks in these domains. The ResNet architecture has also inspired many variations and extensions, such as DenseNets, which use dense connections between layers, and Wide ResNets, which use wider convolutional filters to improve performance \cite{shafiq2022deep}.

\subsubsection{Inception}
Inception is a family of deep convolutional neural network architectures developed by Google researchers for image recognition and classification tasks. The Inception family of models includes Inception V1, V2, V3, V4, and the latest version, Inception-ResNet-v2 \cite{chollet2017xception}.

The fundamental concept underlying the Inception architecture is to employ parallel convolutional layers with various filter sizes. This strategy enables the network to capture features across multiple scales simultaneously. This approach helps to capture both local and global patterns in the image, which leads to better accuracy in image classification tasks.

Inception V1, commonly recognized as GoogLeNet, emerged victorious in the ImageNet Large Scale Visual Recognition Challenge (ILSVRC) within the same year. It encompasses several inception modules, strategically interconnected to forge a profound neural network. The incorporation of these inception modules effectively curtails the computational burden of the network while upholding remarkable accuracy levels. \cite{haritha2020prediction}.

Inception V2 further improved the Inception architecture by introducing several enhancements, including the use of batch normalization, which speeds up the training process and improves the accuracy of the network \cite{szegedy2016rethinking}.

Inception V3  is an extension of the Inception family of models and is designed to achieve better accuracy and efficiency than its predecessors. It includes several enhancements, such as the use of factorized convolutions, which reduce the number of parameters in the network and makes it more efficient \cite{dong2020inception}.

Inception V4 is a more complex version of the Inception architecture that achieves state-of-the-art performance on several image recognition tasks. It includes several advanced techniques, such as residual connections, which allow for better feature reuse across layers \cite{chen2022feature}.

Inception-ResNet-v2 is the latest version of the Inception architecture and combines the Inception and ResNet architectures to achieve even better performance. It includes residual connections and inception modules with factorized convolutions, which make it both accurate and efficient \cite{szegedy2017inception}.

The Inception family of models has become a popular choice for image recognition and classification tasks due to its high accuracy and efficiency. The use of multiple filter sizes and parallel convolutional layers in the inception modules allows the network to capture both local and global patterns in the image, which is essential for accurate image recognition.

\subsubsection{DenseNet}
The architecture of DenseNet revolves around the concept of densely interlinking layers. In this approach, each layer obtains inputs from every preceding layer and transmits its output to all subsequent layers. This stands in contrast to conventional convolutional neural networks (CNNs), where a layer only receives the output of the previous layer. The notion of dense connectivity facilitates every layer's \cite{zhu2017densenet}.

DenseNet is constructed using several dense blocks, each of which comprises multiple convolutional layers possessing an identical count of output channels. The input to the first layer in a dense block is the output of the preceding block. Within each dense block, the output of each layer is concatenated with the outputs of all previous layers and fed as input to the next layer. This results in dense connectivity between layers within a block.

To diminish both the spatial dimensions and the quantity of channels within the feature maps, DenseNet implements transition layers. Each transition layer incorporates a batch normalization layer, followed by a 1x1 convolutional layer and a subsequent 2x2 average pooling layer. The role of the 1x1 convolutional layer is to curtail the number of channels, while the 2x2 average pooling layer reduces the spatial dimensions by a factor of 2. These transition layers are employed after every dense block, except for the final one.

The outcome of the concluding dense block is directed to a global average pooling layer, which computes the average value of each feature map across its spatial extents. The resultant feature vector is subsequently transmitted to a fully connected layer for the purpose of classification. 

DenseNet has consistently demonstrated leading-edge performance across diverse computer vision tasks, encompassing image classification, object detection, and semantic segmentation. Its dense connectivity allows for efficient information flow and encourages feature reuse, which results in high accuracy and reduced training time. However, DenseNet requires more memory than traditional CNNs due to its dense connectivity, which can be a limitation in some applications \cite{zhang2019multiple}.

\subsection{Deep Learning Applications in Diagnosis and Treatment}
In this paper\cite{chen2020epidemiological}, the authors investigates the various aspects of COVID-19 patients, using a retrospective, single-center approach at Jinyintan Hospital in Wuhan. The study scrutinized epidemiological data associated with exposure to virus epicenters, encompassing both short and long-term durations. This analysis covered signs and symptoms, laboratory outcomes, CT scans, and clinical results. Even though the study did not directly focus on predicting COVID-19, it provided valuable insights into the clinical outcomes of the disease. The research highlighted the importance of considering various factors, such as clinical symptoms and laboratory results, in conjunction with imaging techniques such as CT scans, to accurately diagnose and treat COVID-19 patients. The discoveries from this study hold the potential to enhance the management and care of individuals with COVID-19, offering insights that could aid in treatment strategies. Furthermore, these findings may also serve as a valuable point of reference for forthcoming research endeavors in this domain.

In this paper \cite{borkowski2020using}, the authors discussed the potential of deep learning (DL) algorithms in complementing traditional radiographic interpretation for COVID-19 diagnosis from chest X-rays. They emphasize that DL can aid in various ways, including detecting infection, prioritizing patient treatment, and expediting diagnosis (Borkowski et al., 2020). DL algorithms are used to classify and analyze chest images to diagnose COVID-19 from medical images. The procedure entails extracting features from images using the convolutional neural network (CNN) method. Subsequently, these extracted features are utilized in an artificial neural network (ANN) for the purposes of classification and detection.

In this paper \cite{wang2021deep}, the authors' primary goal was to detect visual changes in CT images of COVID-19 patients in China, leveraging deep learning techniques. By extracting distinctive graphical attributes from CT scans, the researchers established an alternative diagnostic approach that demonstrated the feasibility of AI for precise COVID-19 prediction. The team amassed CT images from individuals with confirmed COVID-19 cases, as well as those afflicted with pneumonia. It's important to highlight that this study diverges from our own research, where we employ clinical features and laboratory outcomes for prediction. Nonetheless, Wang et al.'s research holds significance in showcasing the potential of combining CT scans and AI for COVID-19 diagnosis.

In this paper \cite{wang2020clinical}, the authors undertook an extensive investigation, offering an intricate analysis that encompassed epidemiological, demographic, clinical, laboratory, radiological, and treatment-related information sourced from Zhongnan Hospital in Wuhan, China. The primary objective of this study was to meticulously monitor the advancement of COVID-19 infections and contribute to a more profound comprehension of the disease's nature. The authors analyzed and documented this data, which can be useful in predicting COVID-19 in models. Particularly, their detailed analysis of radiological and treatment data provides valuable insights that can aid in the development and improvement of COVID-19 prediction models.

In this paper \cite{maghded2020novel}, the authors introduced an innovative AI framework designed for the identification of COVID-19 through the utilization of built-in sensors within smartphones. This framework harmoniously combines data obtained from diverse sensors to forecast the extent of pneumonia severity and the probability of COVID-19 infection. By relying on multiple readings from different sensors, the proposed framework is designed to detect symptoms associated with COVID-19. The CT scan images uploaded by the patients serve as a crucial element in predicting the presence of COVID-19. The potential of this framework to enable early detection of COVID-19 and improve patient outcomes underscores the importance of leveraging AI technologies in the fight against this global pandemic.

In this paper \cite{narin2021automatic}, the authors proposed an alternative diagnosis option for COVID-19 using an automatic detection system. The research showcases three distinct models based on convolutional neural networks (CNNs): ResNet50, InceptionV3, and Inception-ResNetV2. These models are designed to identify cases of coronavirus pneumonia in patients by analyzing chest X-ray radiographs. The proposed models have demonstrated high accuracy in detecting COVID-19 infections in patients. The authors have compared the classification performance accuracy of the three CNN models and discussed their strengths and weaknesses. The research proposes that the newly developed automatic detection system could function as a valuable and non-intrusive diagnostic tool for accurately diagnosing COVID-19. This system has the potential to aid healthcare professionals in making well-informed decisions regarding patient care. By doing so, this study adds to the increasing body of knowledge regarding the use of artificial intelligence and machine learning in the battle against COVID-19.

In this paper \cite{yan2020prediction}, the authors proposed a new approach to predicting mortality risk in patients using a three-indices-based model. The prognostic prediction model they devised is constructed using the XGBoost machine learning algorithm, known for its ability to straightforwardly and effectively gauge the risk of mortality in patients. While this study is centered around foreseeing mortality risk, it distinctly varies from our own research, which exclusively depends on clinical observations to predict COVID-19 outcomes. By utilizing machine learning algorithms, the authors in this study provide insights into the potential of these algorithms in predicting patient outcomes, which could have important implications in clinical decision-making. However, it is important to note that our study focuses on a different aspect of COVID-19 prediction, and thus, our findings are complementary to this study.

In this paper \cite{ardabili2020covid}, the authors conduct a comprehensive comparative analysis of various machine learning models aimed at predicting the outbreak of COVID-19 across different countries. Their study effectively underscores the capability of machine learning models in forecasting the progression of COVID-19 and its potential spread. However, their analysis is solely based on the outbreak of cases in various countries. In contrast, our work focuses on predicting the disease based on clinical information. We utilize machine learning algorithms to predict COVID-19 severity and mortality in patients, by analyzing their clinical features such as age, symptoms, and medical history. Our work aims to assist healthcare professionals in making informed decisions and improving patient outcomes.

In this paper \cite{purushotham2018benchmarking}, the authors conducted a benchmarking assessment encompassing multiple machine learning algorithms, deep learning algorithms, and ICU scoring systems. These evaluations spanned various clinical prediction tasks and were conducted using publicly accessible clinical datasets. However, their research focuses on COVID-19 patient information exclusively. While this study provides valuable insights into the performance of different algorithms on various clinical prediction tasks, it is not directly relevant to our research objectives. By contrast, their work aims to develop a model that can predict the severity of COVID-19 based on clinical information, which is a critical need in managing the ongoing pandemic.

The use of AI in predicting patient outcomes and aiding healthcare providers in patient triage has become increasingly important during the COVID-19 pandemic. In this paper \cite{alafif2021machine}, the authors presented the potential of AI in predicting the possibility of death by analyzing previous patient data. In situations where there is a surge of COVID-19 cases and healthcare providers are overwhelmed, the triage process becomes crucial. By sorting patients based on their need for immediate medical attention, healthcare providers can prioritize care and potentially save more lives. The use of a deep learning-based algorithm in analyzing medical images has shown promise in aiding the triage process. Such algorithms can quickly analyze large amounts of data to predict patient outcomes and prioritize treatment, ultimately leading to better patient care.

In this paper \cite{jiang2020role}, the authors conducted a study on the use of CT scans in COVID-19 diagnosis, monitoring, and follow-up. In this study, they classified COVID-19 manifestations into four categories based on lesion extent and severity observed in CT scan analysis: early, advanced, critical, and complicated. This classification allows for a better understanding of disease progression and severity. An AI-enabled chest imaging system can potentially assist in tracking the progression of COVID-19 and detecting pulmonary complications quickly.

\subsection{Remarks and Conclusions}
The papers discussed above have emphasized the valuable role of machine learning and artificial intelligence in addressing various aspects of COVID-19, including detection, prediction, and monitoring. These studies have shown promise in predicting the spread of the virus, diagnosing COVID-19 from medical images like chest X-rays and CT scans, and assessing the mortality risk in patients. Nevertheless, it is essential to acknowledge the limitations present in these studies.

A notable limitation across many of these investigations is the reliance on relatively small sample sizes and the use of data from single hospitals or countries. Such constraints can raise concerns about the generalizability of the findings to broader populations. Additionally, the quality of the data employed in these studies has not been consistent, and the validation of some machine learning models on independent datasets has been lacking.

Given these limitations, it is prudent to exercise caution when contemplating the implementation of these machine learning models in clinical practice. While the potential benefits are clear, ensuring the reliability and effectiveness of these models across diverse scenarios is of paramount importance. Therefore, further research is imperative to validate these models on larger and more diverse datasets. Such efforts will be instrumental in harnessing the full potential of machine learning and artificial intelligence against COVID-19.
\newpage

\clearpage
\section{System Architecture}
This chapter involves data collection, data preprocessing, and proposed mythologies designed for the development of a Deep Learning model to predict the severity of COVID-19 in patients.
\subsection{Data Collection}

\subsubsection{Extensive COVID-19 X-Ray and CT Chest Images Dataset}
The first dataset collected from Menoufia University which consisting of both X-ray and CT images with Non-COVID and COVID cases \cite{el2020extensive}. The dataset has been augmented with various techniques to generate around 17099 X-ray and CT images. The dataset is divided into two main folders, one for X-ray images and the other for CT images, with sub-folders for Non-COVID and COVID cases. The X-ray folder contains 5500 Non-COVID and 4044 COVID images, while the CT folder has 2628 Non-COVID and 5427 COVID images. The dataset is available on Mendeley for further use and analysis.

\subsubsection{SARS-COV-2 CT-Scan Dataset}
The second dataset was collected from real patients who were admitted to hospitals located in Sao Paulo, Brazil \cite{soares2020sars}. This dataset included a total of 2482 CT scans that were obtained as part of routine diagnostic procedures for patients exhibiting symptoms of respiratory illnesses or suspected of being infected with SARS-CoV-2. Of the 2482 CT scans in this dataset, 1252 were obtained from patients who had confirmed SARS-CoV-2 infection (COVID-19) and exhibited symptoms associated with the disease. The remaining 1230 scans were obtained from patients who tested negative for SARS-CoV-2 infection. All of the CT scans were performed using a variety of CT scanners and protocols that were available in the different hospitals, reflecting real-world clinical practice. The scans were processed and labeled as either positive or negative for SARS-CoV-2 infection based on radiological findings and the patient's laboratory results.

\subsubsection{COVID-QU-Ex Dataset}
The third dataset is collected from a group of researchers from Qatar University has recently developed a comprehensive dataset called COVID-QU-Ex \cite{tahir2021covid, rahman2021exploring, degerli2021covid, chowdhury2020can}. This dataset comprises a total of 33,920 chest X-ray images, which includes 11,956 images of COVID-19 cases, 11,263 images of non-COVID infections (caused by viral or bacterial pneumonia), and 10,701 images of normal cases. Moreover, this dataset provides ground-truth lung segmentation masks for the entire dataset, which is the largest dataset ever created for lung mask segmentation.

\subsection{Data Loading and Preprocessing}
In the course of this research, three distinct datasets comprising chest X-ray and CT images were amalgamated to form a unified dataset for the purpose of COVID-19 classification. Following the merging of these datasets, we proceeded to categorize the images based on their diagnostic information. The categorization was conducted into two primary groups, distinguishing between CT images and X-ray images: \newpage

\textbf{CT Scan Images:} This category includes all the CT scans from the combined datasets. The CT images within this group represent both COVID-19-positive cases and Non-COVID cases, providing a diverse set of CT scans for analysis.

\textbf{X-ray Images:} In this category, we grouped together all the X-ray images obtained from the combined datasets. These X-ray images encompass both COVID-19-positive cases and Non-COVID cases and form the basis for X-ray-based COVID-19 classification.

This approach ensured the creation of two distinct subsets within the unified dataset, one comprising CT images and the other comprising X-ray images. These subsets facilitated focused investigations and analyses related to COVID-19 classification, enabling the development and evaluation of machine learning models tailored to the respective imaging modalities.
\subsubsection{Data Loading}
This step involves loading and organizing the images into separate lists based on their respective categories: COVID and non-COVID.
First, the code sets the working directory to the location of the dataset. This ensures that the subsequent file operations are performed in the correct directory.
Then, two empty lists, covid images, and not covid images, are created to store the CT scan images of COVID and non-COVID cases, respectively.
The next two loops iterate over the files in the 'COVID' and 'NORMAL' directories. For each file with the '.png' extension, an image object is created using the Image.open() function from the Python Imaging Library (PIL). The os.path.join() function is used to construct the file path by joining the directory name and file name.
Finally, the loaded image objects are appended to their respective lists, covid images and not covid images, based on the directory they were retrieved from.

\subsubsection{Data Preprocessing}
This step is part of the image preprocessing for the covid-19 images datasets. It involves applying a series of operations to each image to prepare it for further analysis or modeling.

The preprocess image function takes an image object as input. Here is a breakdown of the operations performed inside the function:
\begin{enumerate}
\item Resize: The image is resized to a specified size, denoted by IMG SIZE. This operation ensures that all images have the same dimensions for consistency during analysis.
\item Convert to grayscale: The image is converted to grayscale using the 'L' mode. This operation reduces the image's color channels to a single channel, which simplifies subsequent calculations.
\item Convert to array: The image is converted to a NumPy array. This conversion allows for efficient manipulation and processing of the image data.
\item Reshape: The array is reshaped to match the desired dimensions, which include the resized dimensions specified by IMG SIZE and an additional dimension of 1. This is necessary to conform to the expected input shape of certain machine learning models.
\item Normalize: The pixel values in the array are scaled to a range between 0 and 1 by dividing each value by 255. This normalization step ensures that the image data is in a consistent range for effective model training.
\end{enumerate}
The preprocessed images are then stored in separate lists: covid images processed for COVID images and not covid images processed for non-COVID images. The preprocessing steps are applied to each image in the original covid images and not covid images lists using a loop, and the preprocessed image arrays are appended to the respective processed image lists. 

\subsection{Methodologies}
\subsubsection{Convolutional Neural Network(CNN)}
The model commences with the initial convolutional block, comprising a Conv2D layer equipped with 32 filters. Each of these filters scans the input image using a kernel of size (3, 3) and applies a Rectified Linear Unit (ReLU) activation function, thereby introducing non-linearity into the process. This permits the model to grasp and extract diverse low-level features, such as edges and textures. The use of padding='same' ensures that the resultant feature maps maintain the same spatial dimensions as the input. Following this, the feature maps undergo downsampling through a MaxPooling2D layer featuring a pool size of (2, 2). This operation trims down the spatial dimensions while retaining the most essential information.

In the second convolutional block, a similar structure is repeated. This block introduces a Conv2D layer with 64 filters, followed by a MaxPooling2D layer. These additional filters allow the model to capture more complex features and patterns in the input images.

The third convolutional block further enhances the model's feature extraction capabilities. It includes a Conv2D layer with 128 filters, followed by a MaxPooling2D layer. By increasing the number of filters, the model becomes more capable of capturing higher-level features and representations in the input images.

To increase the model's capacity for learning intricate features, two additional Conv2D layers with 256 filters each are included. These layers further refine the learned representations by applying multiple convolutional operations and ReLU activations. Another MaxPooling2D layer is then applied to downsample the feature maps.

The output feature maps are flattened into a 1-dimensional vector following the convolutional layers. This flattening process prepares the data for the subsequent fully connected layers.

Following the flattening process, two Dense layers are incorporated. The initial Dense layer comprises 32 units and employs a Rectified Linear Unit (ReLU) activation function. This layer introduces non-linearity, facilitating the creation of intricate feature combinations. Subsequently, the concluding Dense layer features a solitary unit accompanied by a sigmoid activation function. This configuration is apt for binary classification undertakings. This last layer provides the model's output, producing a probability value between 0 and 1, representing the likelihood of the input image belonging to a particular class. \newpage

Below is the code,
\lstset{
  basicstyle=\ttfamily,
  columns=fullflexible,
  frame=single,
  breaklines=true,
  postbreak=\mbox{\textcolor{red}{$\hookrightarrow$}\space},
}
\begin{lstlisting}[language=python]  
model = Sequential()

model.add(Conv2D(filters=32, kernel_size=(3, 3), activation='relu', input_shape=(*IMG_SIZE, 1), padding='same'))
model.add(MaxPooling2D(pool_size=(2, 2)))

model.add(Conv2D(filters=64, kernel_size=(3, 3), activation='relu', padding='same'))
model.add(MaxPooling2D(pool_size=(2, 2)))

model.add(Conv2D(filters=128, kernel_size=(3, 3), activation='relu', padding='same'))
model.add(MaxPooling2D(pool_size=(2, 2)))

model.add(Conv2D(filters=256, kernel_size=(3, 3), activation='relu', padding='same'))
model.add(Conv2D(filters=256, kernel_size=(3, 3), activation='relu', padding='same'))
model.add(MaxPooling2D(pool_size=(2, 2)))

model.add(Flatten())
model.add(Dense(32, activation='relu'))
model.add(Dense(1, activation='sigmoid'))
\end{lstlisting}

\subsubsection{Densely Connected Convolutional Networks (DenseNet)}
\textbf{DenseNet121} \newline
\newline
In this section, we explain how we implement the DenseNet121 architecture. DenseNet121 is a convolutional neural network (CNN) model pre-trained on a large dataset and is commonly used for image classification tasks.

DenseNet121 is another variant of the DenseNet architecture proposed by Huang et al \cite{huang2017densely}. It follows a similar principle of dense connectivity and feature reuse but with a smaller number of layers compared to DenseNet169 and DenseNet201.

Within DenseNet121, the framework is structured around numerous dense blocks. Each of these dense blocks comprises an assembly of convolutional layers, bolstered by batch normalization and Rectified Linear Unit (ReLU) activation functions. The unique characteristic of dense connectivity is attained by amalgamating the feature maps originating from all prior layers and funneling them into the present layer as input. This interlacing of features across various stages of abstraction empowers the network to amalgamate attributes effectively, stimulating feature reutilization, and facilitating the seamless passage of gradients across the network.

Similar to DenseNet169, DenseNet121 also incorporates transition blocks between dense blocks. The transition blocks are composed of three essential components: a batch normalization layer, a 1x1 convolutional layer, and a 2x2 average pooling layer. These transition blocks serve the vital function of managing spatial dimensions and diminishing the count of feature maps. This strategic adjustment contributes to enhancing computational efficiency and reducing the memory usage of the model.

By connecting each layer to every other layer within the dense blocks, DenseNet121 effectively addresses the vanishing gradient problem that commonly occurs in deep neural networks. The dense connectivity allows the gradients to flow directly from the output layer back to the earlier layers, enabling efficient training and improving the model's ability to learn complex patterns and representations from the input data.

In the below code you can see, the DenseNet121 model is used from the Keras library. The model begins by loading a pre-trained model without its top layers. To ensure that the weights of the base model are not altered during training, these base model layers are frozen. Supplementary layers are then introduced on top of the base model. These additional layers encompass a flattened layer, followed by a fully connected Dense layer that utilizes the Rectified Linear Unit (ReLU) activation function. Concluding this setup is another Dense layer equipped with a sigmoid activation function, designed for tasks related to binary classification. \newline
Below is the code,
\lstset{
  basicstyle=\ttfamily,
  columns=fullflexible,
  frame=single,
  breaklines=true,
  postbreak=\mbox{\textcolor{red}{$\hookrightarrow$}\space},
}
\begin{lstlisting}[language=python]  
base_model = DenseNet121(include_top=False, input_shape=(*IMG_SIZE, 3))

# Freeze the base model layers
base_model.trainable = False

# Create the model architecture by adding the base model and additional layers
model = Sequential()
model.add(base_model)
model.add(Flatten())
model.add(Dense(32, activation='relu'))
model.add(Dropout(0.2))
model.add(Dense(1, activation='sigmoid'))
\end{lstlisting}

\textbf{DenseNet169} \newline
\newline
In this section, we explain how we implement DenseNet169 architecture. DenseNet169 is a convolutional neural network (CNN) model that has been pre-trained on a large dataset and is commonly used for image classification tasks.

DenseNet169 is a robust architecture of deep convolutional neural networks introduced by Huang et al. in their paper titled "Densely Connected Convolutional Networks."\cite{huang2017densely}. It is an extension of the DenseNet architecture, which aims to address the vanishing gradient problem and promote feature reuse by introducing dense connections between layers.

The fundamental concept underpinning DenseNet169 is the interconnection of each layer with every other layer in a sequential manner. This dense connectivity structure empowers the network to tap into the feature maps of all prior layers, which not only eases the flow of gradients and information propagation but also fosters the reutilization of features. Each layer benefits from input sourced from all preceding layers, enabling it to directly access and integrate information from diverse levels of abstraction.

Within the DenseNet169 framework, the structure encompasses numerous dense blocks. These dense blocks are constructed from multiple convolutional layers, complemented by batch normalization and Rectified Linear Unit (ReLU) activation functions. Notably, each dense block is characterized by the concatenation of feature maps from all prior layers. These concatenated feature maps are transmitted as input to the ongoing layer within the block. This interlinking mechanism, termed dense connectivity, guarantees that the network possesses an extensive array of features at every layer.

Additionally, DenseNet169 incorporates transition blocks between dense blocks to control the spatial dimensions and reduce the number of feature maps. These transition blocks consist of a batch normalization layer, a 1x1 convolutional layer, and a 2x2 average pooling layer. They help in reducing the spatial dimensions while preserving the information richness.

In the implementation, we use the DenseNet169 model from the Keras library. The procedure entails loading a pre-trained model while excluding the top layers (fully connected). Subsequently, the base model layers are frozen to inhibit any alteration to their weights during training. On top of the base model, we introduce our own layers. Ultimately, the model is compiled by specifying an appropriate loss function, optimizer, and metrics.

By training the DenseNet169 model on a specific task, such as image classification or object detection, it can learn to extract and leverage hierarchical features from the input data, leading to improved performance and accuracy compared to traditional convolutional neural network architectures.\newline
Below is the code, 

\lstset{
  basicstyle=\ttfamily,
  columns=fullflexible,
  frame=single,
  breaklines=true,
  postbreak=\mbox{\textcolor{red}{$\hookrightarrow$}\space},
}
\begin{lstlisting}[language=python]  
base_model = DenseNet169(include_top=False, input_shape=(*IMG_SIZE, 3))

# Freeze the base model layers
base_model.trainable = False

# Create the model architecture by adding the base model and additional layers
model = Sequential()
model.add(base_model)
model.add(Flatten())
model.add(Dense(32, activation='relu'))
model.add(Dropout(0.2))
model.add(Dense(1, activation='sigmoid'))
\end{lstlisting}

\textbf{DenseNet201} \newline
\newline
DenseNet201 is an extension of the DenseNet architecture that builds on the concepts introduced in DenseNet121 and DenseNet169. It is a deeper and more complex network designed to capture even more intricate patterns and representations in the input data \cite{huang2017densely}.

In DenseNet201, the core idea of dense connectivity remains the same. The network structure comprises dense blocks, each of which encompasses multiple convolutional layers. However, compared to DenseNet121 and DenseNet169, DenseNet201 has a significantly larger number of layers, allowing it to learn more intricate features and capture more hierarchical information.

Similar to other DenseNet variants, DenseNet201 incorporates transition blocks between dense blocks. These transition blocks help control the spatial dimensions and reduce the number of feature maps, ensuring computational efficiency and memory usage optimization.

The dense connectivity in DenseNet201 allows feature maps from all preceding layers to be directly connected to the current layer, promoting feature reuse and enabling effective information flow throughout the network. This dense connectivity contributes to alleviating the vanishing gradient problem and facilitates better gradient propagation, enabling the network to learn more efficiently.

Capitalizing on its deep and densely interconnected design, DenseNet201 showcases robust representational capabilities, rendering it well-suited for an array of computer vision endeavors. These encompass image classification, object detection, and semantic segmentation. Its adeptness at capturing intricate patterns and harnessing feature reutilization positions it to attain cutting-edge performance on demanding datasets.

In terms of implementation, we use the pre-trained model without the top layers, freezing the base model layers, and adding additional layers for specific task requirements, such as classification or segmentation. This allows for fine-tuning or transfer learning, where the pre-trained weights capture general image features that can be adapted to a specific task with a smaller dataset. \newline
Below is the code,

\lstset{
  basicstyle=\ttfamily,
  columns=fullflexible,
  frame=single,
  breaklines=true,
  postbreak=\mbox{\textcolor{red}{$\hookrightarrow$}\space},
}
\begin{lstlisting}[language=python]  
base_model = DenseNet201(include_top=False, input_shape=(*IMG_SIZE, 3))

# Freeze the base model layers
base_model.trainable = False

# Create the model architecture by adding the base model and additional layers
model = Sequential()
model.add(base_model)
model.add(Flatten())
model.add(Dense(32, activation='relu'))
model.add(Dense(1, activation='sigmoid'))
\end{lstlisting}

\subsubsection{Visual Geometry Group(VGG)}
\textbf{VGG16} \newline
\newline
VGG16 is a convolutional neural network architecture formulated by the Visual Geometry Group (VGG) at the University of Oxford (Simonyan and Zisserman, 2014). This architecture has garnered attention for its straightforwardness and efficiency in addressing image classification endeavors.The "VGG" in VGG16 stands for the research group's name \cite{simonyan2014very}.

VGG16 comprises a grand total of 16 layers, encompassing 13 convolutional layers and 3 fully connected layers. This architecture adheres to a sequence of stacking numerous convolutional layers, each with a receptive field of 3x3. The "same" padding technique is employed to retain the spatial dimensions. Following the convolutional layers are max pooling layers with a window size of 2x2 and a stride of 2. These pooling layers gradually diminish the spatial dimensions of the feature maps.

As the network delves deeper, the convolutional layers exhibit a progressive augmentation in the count of filters. The initial layers specialize in capturing rudimentary features like edges and textures, whereas the deeper layers are tasked with apprehending more intricate and abstract attributes. VGG16 utilizes a deeper network compared to previous architectures like AlexNet, which allows it to learn richer representations.

After the convolutional layers, VGG16 adds three fully connected layers. The feature maps from the last convolutional layer are flattened and fed into these dense layers for classification. The fully connected layers gradually reduce the dimensionality until reaching the final output layer.

VGG16 has been widely used as a baseline architecture for image classification tasks and has achieved competitive performance on benchmark datasets. Its simplicity and effectiveness have made it a popular choice for transfer learning, where pre-trained VGG16 models trained on large-scale datasets are utilized as feature extractors or fine-tuned for specific tasks.\newline
Below is the code,

\lstset{
  basicstyle=\ttfamily,
  columns=fullflexible,
  frame=single,
  breaklines=true,
  postbreak=\mbox{\textcolor{red}{$\hookrightarrow$}\space},
}
\begin{lstlisting}[language=python]  
from tensorflow.keras.applications import VGG16
from tensorflow.keras.models import Sequential
from tensorflow.keras.layers import Flatten, Dense

# Load the VGG16 model without the top (fully connected) layers
base_model = VGG16(weights='imagenet', include_top=False, input_shape=(*IMG_SIZE, 3))

# Freeze the base model layers
base_model.trainable = False

# Create the model architecture by adding the base model and additional layers
model = Sequential()
model.add(base_model)
model.add(Flatten())
model.add(Dense(256, activation='relu'))
model.add(Dense(1, activation='sigmoid'))  
\end{lstlisting}

In this code, we first load the pre-trained VGG16 model from the Keras library using the VGG16 function. We specify the weights parameter as 'imagenet' to use the weights pre-trained on the ImageNet dataset. We also set include top to False to exclude the top (fully connected) layers of the model.

Subsequently, we proceed to immobilize the base model layers by setting `base model.trainable` to `False`. This strategic action effectively halts any modification to the weights of the base model during the training process.

We then create the model architecture by adding the base model to a Sequential model and appending additional layers. In this example, we add a Flatten layer to flatten the feature maps, followed by two fully connected (Dense) layers with 4096 units each and ReLU activation. Finally, we add the output layer with Dense and softmax activation, assuming 1000 classes for the ImageNet dataset.

After defining the architecture, we compile the model by specifying the loss function, optimizer, and evaluation metrics. In this case, we use 'binary crossentropy' as the loss function.

Finally, we train the model by calling the fit method and providing the training data (X train, y train). We also specify the batch size, number of epochs, and optionally the validation data (X test, y test) for monitoring the model's performance during training.
\newline
\newline
\textbf{VGG19} \newline
\newline
VGG19 is a deep convolutional neural network architecture that follows a similar structure as VGG16 but with a deeper network depth \cite{simonyan2014very}. It consists of 19 weight layers, including 16 convolutional layers and 3 fully connected layers. \newline

The key idea behind VGG19 is to use a stack of small-sized (3x3) convolutional filters with a stride of 1 and a small receptive field, followed by max-pooling layers to achieve an effective receptive field. This design pattern of using multiple stacked convolutional layers with small filters helps VGG19 learn rich and expressive feature representations.

To implement VGG19, we use the Pre-trained model from the Keras library using the VGG19 function. The code structure is similar to VGG16, with the appropriate model import and modifications based on the number of classes and other specific requirements of our task. \newline
Below is the code,

\begin{lstlisting}[language=python]  
from tensorflow.keras.applications import VGG19
from tensorflow.keras.models import Sequential
from tensorflow.keras.layers import Flatten, Dense

# Create the VGG19 base model
base_model = VGG19(weights='imagenet', include_top=False, input_shape=(*IMG_SIZE, 3))

# Freeze the base model layers
base_model.trainable = False

# Create the model architecture by adding the base model and additional layers
model = Sequential()
model.add(base_model)
model.add(Flatten())
model.add(Dense(4096, activation='relu'))
model.add(Dense(4096, activation='relu'))
model.add(Dense(1, activation='sigmoid'))
\end{lstlisting}

\subsubsection{Inception}
\textbf{Inception V3} \newline
\newline
The Inception V3 model is a deep convolutional neural network architecture that aims to enhance performance through the use of multiple convolutional filters with different receptive field sizes. By employing a combination of 1x1, 3x3, and 5x5 convolutional filters along with pooling layers, Inception V3 seeks to capture features at various spatial scales efficiently. This architecture reduces the risk of information loss and allows the network to extract both local and global features from the input data \cite{szegedy2016rethinking}.

In the implementation of the Inception V3 model for COVID-19 classification, we utilized the TensorFlow library. The image data was preprocessed using the `ImageDataGenerator` to apply data augmentation techniques, enhancing the model's ability to generalize effectively. We loaded the InceptionV3 model with pretrained weights from ImageNet while excluding the top fully connected layers. The model architecture was extended by adding a Global Average Pooling layer followed by custom fully connected layers. The final layer with a sigmoid activation function facilitated binary classification. To avoid overfitting, we froze the layers of the base model, allowing only the newly added layers to be trained. This approach leveraged the powerful feature extraction capabilities of the Inception V3 model for effective COVID-19 classification on X-ray images. \newline
Below is the code,
\begin{lstlisting}[language=python]  
from tensorflow.keras.applications import InceptionV3

# Load the InceptionV3 model with pretrained weights (excluding the top fully connected layers)
base_model = InceptionV3(include_top=False, input_shape=(*IMG_SIZE, 3), weights='imagenet')

# Add custom fully connected layers for binary classification
x = base_model.output
x = GlobalAveragePooling2D()(x)
x = Dense(128, activation='relu')(x)
predictions = Dense(1, activation='sigmoid')(x)

# Create the final model
model = Model(inputs=base_model.input, outputs=predictions)

# Freeze the layers of the base model
for layer in base_model.layers:
    layer.trainable = False


\end{lstlisting}

\textbf{Inception V4} \newline
\newline
The InceptionV4 architecture, a refinement of its predecessors like InceptionV3, stands out as a sophisticated convolutional neural network designed to tackle complex image recognition tasks. It builds upon the idea of using multiple filters of varying sizes to capture diverse levels of features within the input data. This enables the network to effectively represent intricate patterns and structures, making it particularly suitable for tasks such as classifying medical images, including COVID-19 X-ray scans \cite{szegedy2017inception}.

We undertake the development of the InceptionV4 model from the ground up to categorize COVID-19 X-ray images. InceptionV4 is a sophisticated convolutional neural network architecture acclaimed for its proficiency in image recognition assignments.

The model is constructed using various layers, including convolutional, pooling, and fully connected layers, organized in a modular structure. The key building blocks are "inception blocks," which consist of multiple convolutional paths with different filter sizes. These paths capture different scales of features within the input data and are concatenated to provide a richer representation.

The model begins with a stem module that includes convolutional layers to process the input image. Then, a series of inception blocks are applied to progressively extract hierarchical features. MaxPooling2D layers are utilized to downsample the spatial dimensions and increase the receptive field. Finally, a global average pooling layer and fully connected layers are added for classification.

The `create inceptionv4 model` function defines the complete architecture, including the stem, multiple inception blocks, and the classifier. The model is designed to accept input images with the specified shape and output a sigmoid activation for binary classification.

This custom InceptionV4 model is tailored to the COVID-19 X-ray dataset and aims to capture relevant patterns and features for accurate classification. The implementation from scratch allows for fine-tuning the architecture to the characteristics of the dataset.

It's important to note that implementing a complex architecture like InceptionV4 from scratch requires careful tuning and experimentation to achieve optimal results. This approach provides flexibility in designing a model tailored to the specific problem domain and dataset.\newline
Below is the code,
\begin{lstlisting}[language=python]  
def inception_block(x, filters):
    path1 = Conv2D(filters[0], (1, 1), padding='same', activation='relu')(x)

    path2 = Conv2D(filters[1], (1, 1), padding='same', activation='relu')(x)
    path2 = Conv2D(filters[2], (3, 3), padding='same', activation='relu')(path2)

    path3 = Conv2D(filters[3], (1, 1), padding='same', activation='relu')(x)
    path3 = Conv2D(filters[4], (3, 3), padding='same', activation='relu')(path3)
    path3 = Conv2D(filters[5], (3, 3), padding='same', activation='relu')(path3)

    path4 = AveragePooling2D((3, 3), strides=(1, 1), padding='same')(x)
    path4 = Conv2D(filters[6], (1, 1), padding='same', activation='relu')(path4)

    return Concatenate()([path1, path2, path3, path4])

def create_inceptionv4_model(input_shape=(*IMG_Size, 3), num_classes=1):
    inputs = Input(shape=input_shape)

    # Stem
    x = Conv2D(32, (3, 3), padding='same', activation='relu')(inputs)
    x = Conv2D(32, (3, 3), padding='same', activation='relu')(x)
    x = Conv2D(64, (3, 3), padding='same', activation='relu')(x)

    # Inception blocks
    x = inception_block(x, [64, 96, 128, 16, 32, 32, 32])
    x = inception_block(x, [128, 128, 192, 32, 96, 64, 64])
    x = MaxPooling2D(pool_size=(3, 3), strides=(2, 2), padding='same')(x)

    # Add more inception blocks as needed
    # Classifier
    x = Flatten()(x)
    x = Dense(256, activation='relu')(x)
    x = Dense(num_classes, activation='sigmoid')(x)
    model = Model(inputs=inputs, outputs=x)
    return model

# Create the InceptionV4 model
model = create_inceptionv4_model()
\end{lstlisting}

\subsubsection{AlexNet}
AlexNet is an impactful deep convolutional neural network architecture that garnered substantial recognition and prominence following its triumph in the ImageNet Large Scale Visual Recognition Challenge (ILSVRC). This model served as one of the trailblazers, showcasing the efficacy of deep learning in image classification endeavors \cite{krizhevsky2017imagenet}. The architecture of AlexNet consists of eight layers, including five convolutional layers and three fully connected layers.

The convolutional layers in AlexNet extract hierarchical features from input images by applying a series of convolutional filters. These filters learn to detect various patterns and features at different levels of abstraction. The use of multiple convolutional layers allows the network to capture both local and global information in the input images.

The fully connected layers in AlexNet are responsible for high-level reasoning and decision-making. They take the extracted features from the convolutional layers and map them to the desired output classes. The fully connected layers are typically deeper and have a larger number of parameters compared to the convolutional layers.

AlexNet also introduced several key architectural innovations that have influenced subsequent models. It utilized Rectified Linear Units (ReLU) as the activation function, which helps alleviate the vanishing gradient problem and accelerates convergence. Additionally, AlexNet employed techniques such as overlapping pooling, local response normalization, and dropout to reduce overfitting and improve generalization performance.

The provided code snippet outlines the architecture of the AlexNet convolutional neural network model which we implement for COVID-19 classification. It encompasses a series of interconnected layers, each contributing to feature extraction and classification tasks. The network initiates with a pivotal convolutional layer employing 96 filters of size 11x11, followed by a Rectified Linear Unit (ReLU) activation function. Subsequently, a pooling layer reduces spatial dimensions using a 3x3 window, thus extracting salient features.

A subsequent convolutional layer integrates 256 filters of size 5x5, maintaining spatial dimensions through padding. Employing ReLU activation and another pooling layer, the network aims to capture intricate patterns. The architecture continues to unfold with two consecutive convolutional layers, both employing 384 filters of size 3x3, further refining feature extraction. The fifth convolutional layer utilizes 256 filters of size 3x3, contributing to the model's understanding of complex visual features. A final pooling layer diminishes spatial dimensions.

Following the convolutional layers, a flatten layer converts the multidimensional output into a linear format. The model then incorporates two densely connected layers, each consisting of 4096 neurons with ReLU activation. These layers serve as intermediaries, enhancing the understanding of intricate relationships within the data. The model ultimately culminates in an output layer, utilizing a sigmoid activation function for binary classification tasks. Overall, this architecture encapsulates a comprehensive pipeline for feature extraction and learning, demonstrating the essence of the AlexNet model's significant impact in the realm of deep learning. \newpage
Below is the code,

\begin{lstlisting}[language=python]  

# Create the AlexNet model
model = Sequential()

# First Convolutional Layer
model.add(Conv2D(filters=96, kernel_size=(11, 11), strides=(4, 4), activation='relu', input_shape=(*IMG_SIZE, 3), padding='valid'))
model.add(MaxPooling2D(pool_size=(3, 3), strides=(2, 2)))

# Second Convolutional Layer
model.add(Conv2D(filters=256, kernel_size=(5, 5), strides=(1, 1), activation='relu', padding='same'))
model.add(MaxPooling2D(pool_size=(3, 3), strides=(2, 2)))

# Third Convolutional Layer
model.add(Conv2D(filters=384, kernel_size=(3, 3), strides=(1, 1), activation='relu', padding='same'))

# Fourth Convolutional Layer
model.add(Conv2D(filters=384, kernel_size=(3, 3), strides=(1, 1), activation='relu', padding='same'))

# Fifth Convolutional Layer
model.add(Conv2D(filters=256, kernel_size=(3, 3), strides=(1, 1), activation='relu', padding='same'))
model.add(MaxPooling2D(pool_size=(3, 3), strides=(2, 2)))

# Flatten the output from the last convolutional layer
model.add(Flatten())

# Fully Connected Layers
model.add(Dense(4096, activation='relu'))

model.add(Dense(4096, activation='relu'))

# Output layer for binary classification
model.add(Dense(1, activation='sigmoid'))

\end{lstlisting}
\newpage
\subsection{Remarks and Conclusions}
We've harnessed a range of powerful deep-learning models. These models, including DenseNet121, DenseNet169, DenseNet201, VGG16, VGG19, Inception V3, Inception V4, and AlexNet, have been pivotal in our efforts to accurately classify COVID-19 cases from medical images like X-rays and CT scans. Through a combination of detailed architectural insights and practical code implementations, we've tailored and fine-tuned these models to cater specifically to the distinctive features present in COVID-19 imagery. By leveraging the inherent capabilities of these models, we've made significant strides in the development of essential tools for medical professionals and researchers, aiding in the prompt and precise diagnosis of COVID-19. These models stand as a testament to the critical role of deep learning in advancing healthcare and addressing global health challenges.\newline
Deep learning models have revolutionized the field of artificial intelligence and have found wide-ranging applications across various domains, including healthcare. In the context of COVID-19 detection and diagnosis, the utilization of deep learning models has been particularly instrumental. Let's delve further into the significance and characteristics of these models:

DenseNet Models (DenseNet121, DenseNet169, DenseNet201): DenseNet, short for Densely Connected Convolutional Networks, is known for its unique architecture where each layer is connected to every other layer in a feedforward fashion. This dense connectivity allows for better feature reuse and gradient flow, making it well-suited for medical image classification tasks. These models have excelled in extracting complex patterns and features from medical images, enhancing the accuracy of COVID-19 diagnosis.

VGG Models (VGG16, VGG19): The VGG (Visual Geometry Group) models are characterized by their simplicity and uniform architecture. While not as deep as some other architectures, VGG models have demonstrated effectiveness in feature extraction from images, which is crucial in identifying the subtle characteristics of COVID-19 in X-rays and CT scans.

Inception Models (Inception V3, Inception V4): The Inception models, also known as GoogleNet, utilize inception modules with multiple filter sizes. This allows them to capture features at various scales within an image, making them adept at discerning intricate details in medical images. Their adaptability has proven valuable in COVID-19 image classification.

AlexNet: Its pioneering deep convolutional neural network architecture, played a pivotal role in popularizing deep learning. While it may not be as complex as some newer models, it remains relevant due to its ability to handle image classification tasks efficiently. It provides a solid foundation for COVID-19 image analysis.

The fine-tuning and customization of these deep learning models for COVID-19 detection involve adapting the model architectures, optimizing hyperparameters, and training them on COVID-19-specific datasets. This process tailors these models to effectively capture the unique features associated with the disease in medical images, ensuring high accuracy in diagnosis.

The success of these models in the realm of COVID-19 diagnosis highlights the transformative potential of deep learning in healthcare. It showcases how cutting-edge technologies can be harnessed to create essential tools that aid medical professionals and researchers in making rapid and accurate diagnoses, especially in the face of global health challenges like the COVID-19 pandemic. As the field of deep learning continues to advance, we can anticipate even more groundbreaking applications that will further enhance healthcare and improve our ability to combat diseases.
\clearpage
\section{Results and Discussion}
In this chapter, we'll start by explaining the metrics we use to evaluate our models. Then, we'll give a brief explanation of the results from all the models we've implemented. Finally, we'll provide some remarks and draw conclusions based on these results.
\subsection{Performance Metrics Explanation}

In our COVID-19 classification project, we employ several performance metrics to assess the effectiveness of our deep learning models \cite{ferrer2022analysis}. These metrics help us understand the model's classification capabilities and guide us in making informed decisions about model performance.

One fundamental performance evaluation tool used in our project is the confusion matrix. A confusion matrix is a table that is particularly useful in binary classification tasks, such as distinguishing COVID-19 cases from non-COVID-19 cases in medical images. It presents a clear summary of the model's predictions compared to the actual class labels. Table 4.1 displays the structure of the confusion matrix as follows:

\begin{table}[h]
\centering
\caption{Confusion Matrix for COVID-19 Classification}
\begin{tabular}{|c|c|c|}
\hline
\textbf{Actual / Predicted} & \textbf{Positive (COVID-19)} & \textbf{Negative (Non-COVID-19)} \\
\hline
\textbf{Positive (COVID-19)} & True Positives (TP) & False Negatives (FN) \\
\hline
\textbf{Negative (Non-COVID-19)} & False Positives (FP) & True Negatives (TN) \\
\hline
\end{tabular}
\end{table}

\begin{itemize}
  \item \textbf{True Positives (TP):} Number of Cases correctly predicted as positive (COVID-19).
  \item \textbf{False Positives (FP):} Number of Cases incorrectly predicted as positive when they are negative (non-COVID-19) in the context of COVID-19 classification.
  \item \textbf{True Negatives (TN):} Number of Cases correctly predicted as negative (non-COVID-19).
  \item \textbf{False Negatives (FN):} Number of Cases incorrectly predicted as negative when they are positive (COVID-19) in the context of COVID-19 classification.
\end{itemize}

The confusion matrix provides a comprehensive view of the model's performance, allowing us to calculate various performance metrics such as accuracy, precision, recall (sensitivity), specificity, and F1-score. These metrics help us gauge different aspects of the model's performance, such as its ability to correctly classify COVID-19 cases (sensitivity) and its ability to avoid misclassifying non-COVID-19 cases (specificity). By analyzing these metrics and the confusion matrix, we can make informed decisions about model performance and fine-tune our deep learning models to achieve better results in COVID-19 classification.

\begin{itemize}
    \item \textbf{Accuracy}: This metric gauges the comprehensive accuracy of the model's predictions. It is determined by dividing the count of accurate predictions (including both true positives and true negatives) by the total number of instances present in the dataset \cite{mcgwire2001spatially}.
    
    \begin{equation}\label{eq:accuracy}
    \text{Accuracy} = \frac{\text{True Positives} + \text{True Negatives}}{\text{Total Instances}}
    \end{equation}
    
    \item \textbf{Precision}: Precision serves as a metric that hones in on the precision of positive predictions rendered by the model. It quantifies the proportion of true positive predictions against the entire count of positive predictions. Elevated precision implies that the model exercises prudence when making positive predictions, thereby reducing the likelihood of false positives \cite{arias2023confusion}.
    
    \begin{equation}\label{eq:precision}
    \text{Precision} = \frac{\text{True Positives}}{\text{True Positives} + \text{False Positives}}
    \end{equation}
    
    \item \textbf{Recall}: Recall, alternately referred to as True Positive Rate or Sensitivity, assesses the model's proficiency in accurately recognizing positive instances. This metric is computed by dividing the count of true positive predictions by the total number of genuine positive instances (encompassing true positives and false negatives). Heightened recall signifies that the model is skilled at detecting a substantial portion of positive cases \cite{davis2006relationship}.
    
    \begin{equation}\label{eq:recall}
        \text{Recall} = \frac{\text{True Positives}}{\text{True Positives} + \text{False Negatives}}
    \end{equation}
    
    \item \textbf{F1 Score}: The F1 Score represents a balance between precision and recall, delivering a comprehensive assessment of the model's effectiveness. It encompasses the assessment of false positives and false negatives, rendering it especially valuable when handling datasets with imbalanced class distributions. Computed as the harmonic mean of precision and recall, the F1 Score furnishes a singular value that accounts for both dimensions \cite{chicco2020advantages}.
    
    \begin{equation}\label{eq:f1score}
        \text{F1 Score} = 2 \cdot \frac{\text{Precision} \cdot \text{Recall}}{\text{Precision} + \text{Recall}}
    \end{equation}
    
    \item \textbf{Loss}: The loss is a metric that quantifies how often a classification model makes incorrect predictions. It is calculated as the ratio of false positives and false negatives to the total number of instances in the dataset. A lower misclassification rate indicates better model accuracy, while a higher rate suggests a less accurate model \cite{wang2020comprehensive}.

    \begin{equation}\label{eq:misclassification_rate}
    \text{Loss} = \frac{\text{False Positives} + \text{False Negatives}}{\text{Total Instances}}
\end{equation}

\end{itemize}

These metrics provide insights into different aspects of model performance, helping us interpret results and make necessary improvements. By evaluating these metrics, we enhance our COVID-19 classification efforts and ensure the accuracy and reliability of our models.
\newpage

\subsection{CNN}
\subsubsection{Results of CNN on CT Scan Images}
In initial experiment, we employed a Convolutional Neural Network (CNN) architecture for the classification of CT Scan dataset. The CNN architecture we adopted encompasses multiple layers of convolutions, strategically designed to extract intricate features from the input images. Accompanied by pooling layers, which facilitate downsampling while retaining crucial details, this architecture forms the foundation of our model. Our choice of a batch size of 32 and training the model over the course of 20 epochs defined the iterative learning process, where the model was refined through multiple passes over the dataset. Figure 4.1 represents the CNN results on the CT scan images.
\begin{figure}[ht]
    \centering
    \includegraphics[width=0.8\textwidth]{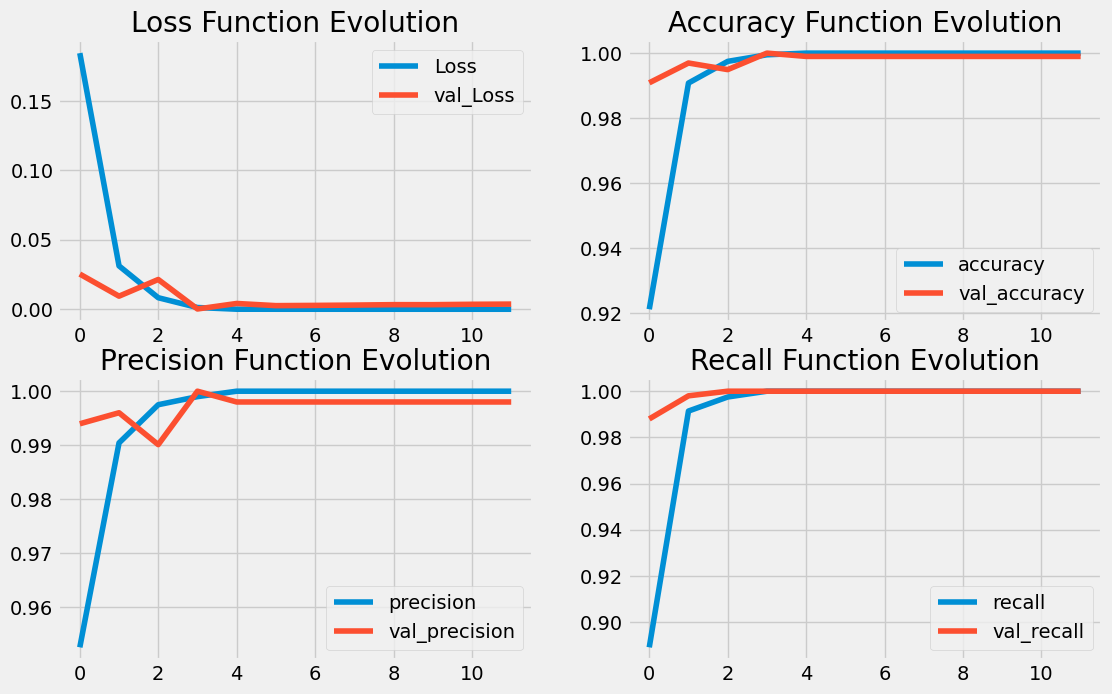}
    \caption{Results of CNN on CT Scan Images}
\end{figure}

Upon scrutinizing the model's performance, we garnered the following insights: Achieving an accuracy of approximately 99.90 percent highlighted the model's exceptional proficiency in aligning predictions with the actual labels. The test precision of around 99.80 percent underscored the model's reliability in generating accurate positive predictions. Moreover, the test loss of 0.0037 indicated a minimal divergence between predictions and actual labels, validating the model's precision. Table 4.2 represents the CNN results on the CT scan images.

\begin{table}[ht]
\centering
\caption{Performance Metrics of CNN on CT Scan Images}
\begin{tabular}{|c|c|c|}
\hline
\textbf{Performance Metrics} & \textbf{Training Result} & \textbf{Testing Result} \\
\hline
Accuracy & 1.0000 & 0.9989 \\
Loss & 0.0001 & 0.0037 \\
Precision & 1.0000 & 0.9980 \\
Recall & 1.0000 & 1.0000 \\
\hline
\end{tabular}
\end{table}

\newpage
\subsubsection{Results of CNN on Chest-X-ray Images}
The same architecture of the CNN model was evaluated on the Chest X-ray Images. During training, the model exhibited impressive capabilities, achieving a training loss of 0.0154. This low loss signifies that the model's predictions closely matched the actual values. Furthermore, the training accuracy reached an exceptional value of 99.50 percent, highlighting the model's proficiency in accurately classifying Chest X-ray images during the training process. Figure 4.2 represents the CNN results on the X-ray Images.

\begin{figure}[ht]
    \centering
    \includegraphics[width=0.8\textwidth]{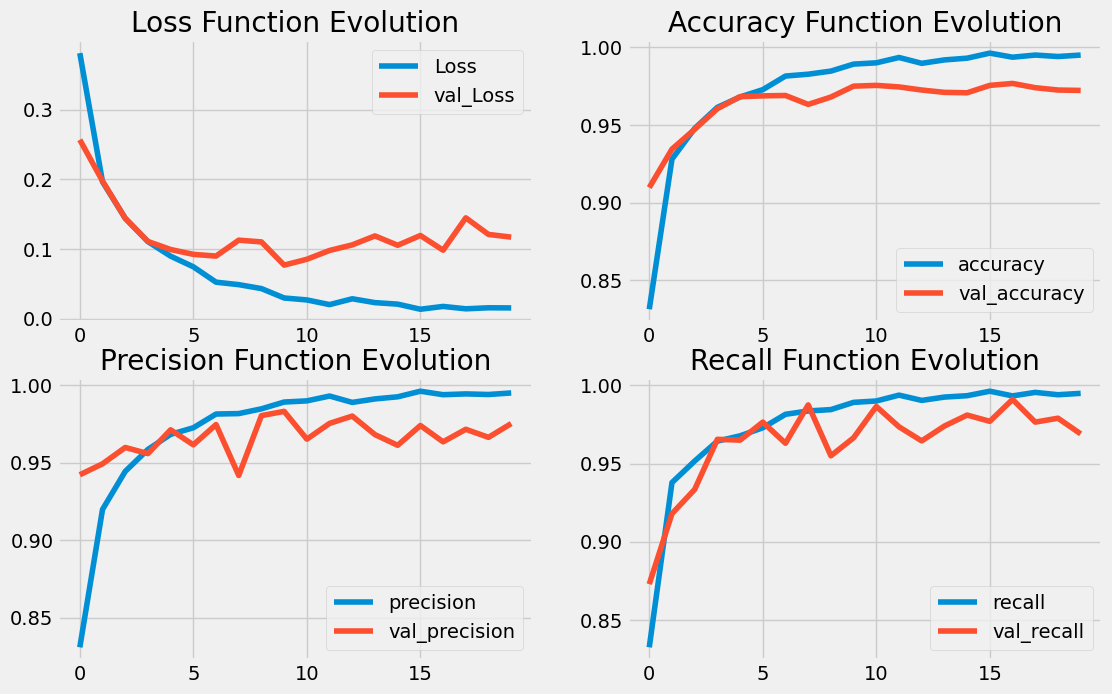}
    \caption{Results of CNN on Chest X-ray Images}
\end{figure}

In terms of precision, the model achieved a precision score of 0.9951 during training. This indicates that the model's positive predictions were highly accurate and reliable. The model's ability to identify true positive instances was also remarkable, with a recall score of 0.9949.

Upon evaluating the model's performance, the results remained consistently impressive. The validation loss was measured at 0.1171, and the validation accuracy achieved a value of 97.22 percent. While the validation accuracy slightly deviated from the training accuracy, the model demonstrated robust generalization to new, unseen data. The precision and recall values on the validation set were measured at 0.9753 and 0.9690, respectively. These scores indicate that the model retained its capability to make accurate predictions on data it had not encountered during training. Table 4.3 represents the CNN results on the X-ray images.

\begin{table}[ht]
\centering
\caption{Performance Metrics of CNN on Chest X-ray Images}
\begin{tabular}{|c|c|c|}
\hline
\textbf{Performance Metrics} & \textbf{Training Result} & \textbf{Testing Result} \\
\hline
Accuracy & 0.9950 & 0.9722 \\
Loss & 0.0154 & 0.1171 \\
Precision & 0.9951 & 0.9753 \\
Recall & 0.9949 & 0.9690 \\
\hline
\end{tabular}
\end{table}

\subsection{DenseNet121}
\subsubsection{Results of DenseNet121 on CT Scan Images}
The DenseNet121 pre-trained model, designed with a 128x128 pixel input size, was employed to analyze COVID-19 CT Scan images. The model's base layers were frozen to retain their learned features. Upon running the model on the CT Scan COVID-19 dataset, the achieved results were as follows. The training loss was measured at 0.0080, indicating that the model's predictions were closely aligned with the actual values. The training accuracy was an impressive 99.74 percent, reflecting the model's proficiency in accurately classifying CT Scan COVID-19 images during the training phase. Figure 4.3 represents the DenseNet121 results on the CT scan images.

\begin{figure}[ht]
    \centering
    \includegraphics[width=0.8\textwidth]{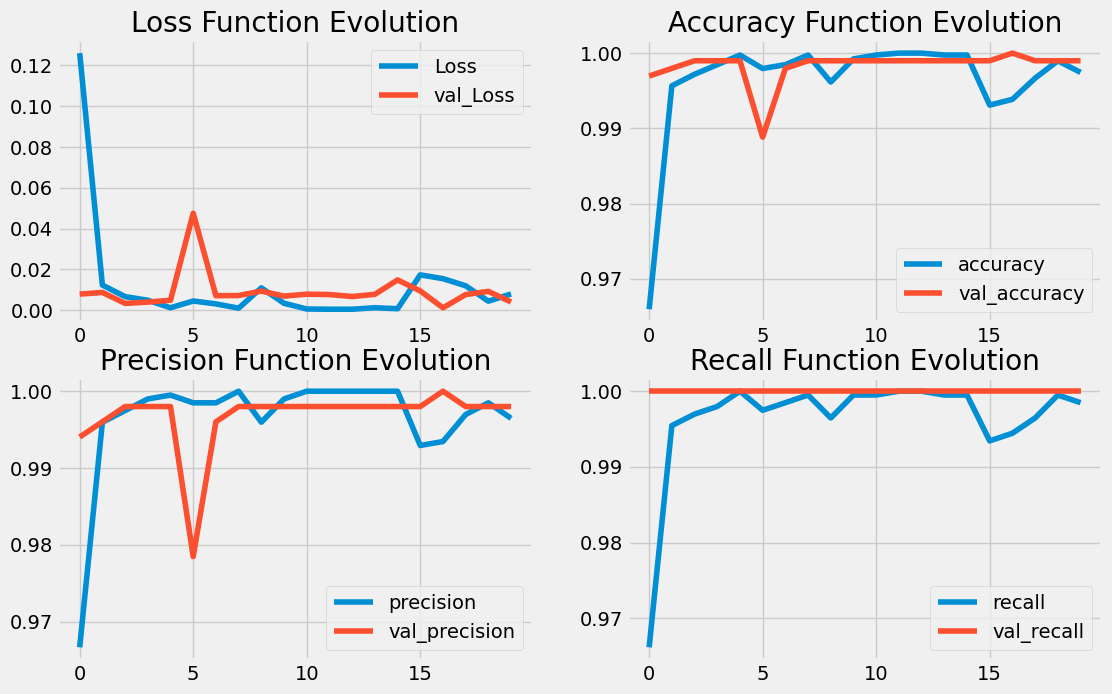}
    \caption{Results of DenseNet121 on CT scan Images}
\end{figure}

The model's precision, a measure of accurate positive predictions, was calculated at 0.9965 during training. This value indicates that the model's positive predictions were highly reliable. Notably, the model demonstrated an excellent recall of 0.9985, signifying its ability to accurately identify true positive instances, further contributing to the model's effectiveness in disease detection.The evaluation of the model's performance yielded equally remarkable outcomes. The validation loss was measured at 0.0041, and the validation accuracy reached an outstanding value of 99.90 percent. The validation precision, indicating accurate positive predictions, was determined as 0.9980, while the validation recall, indicating the ability to identify true positive instances, achieved a perfect score of 1.0000. Table 4.4 represents the DenseNet121 results on the CT scan images.

\begin{table}[ht]
\centering
\caption{Performance Metrics of DenseNet121 on CT Scan Images}
\begin{tabular}{|c|c|c|}
\hline
\textbf{Performance Metrics} & \textbf{Training Result} & \textbf{Testing Result} \\
\hline
Accuracy & 0.9974 & 0.9990 \\
Loss & 0.0080 & 0.0041 \\
Precision & 0.9965 & 0.9980 \\
Recall & 0.9985 & 1.0000 \\
\hline
\end{tabular}
\end{table}

\subsubsection{Results of DenseNet121 on X-ray Images}
The DenseNet121 pretrained model was applied to Covid-19 X-ray images, yielding the following results. The obtained loss value of 0.1863 signifies the discrepancy between predicted and actual values during training. An accuracy of 0.9162 indicates the model's success in accurately predicting outcomes in about 91.62 percent of cases. The model's precision, at 0.9303, reflects its ability to correctly identify true positive cases among all predicted positives. A recall value of 0.8999 signifies the model's effectiveness in identifying actual Covid-19 cases among all true positives. Figure 4.4 represents the DenseNet121 results on the X-ray images.

\begin{figure}[ht]
    \centering
    \includegraphics[width=0.8\textwidth]{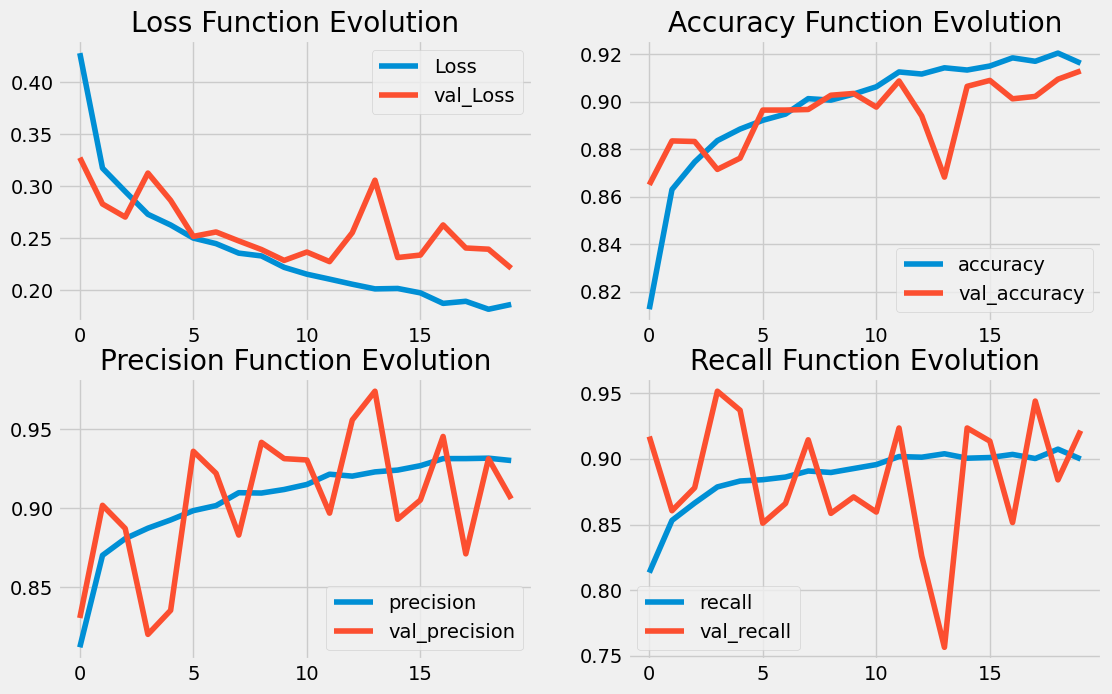}
    \caption{Results of DenseNet121 on X-ray Images}
\end{figure}

In the validation phase, the model demonstrated a validation loss of 0.2211, comparable to its training performance. The validation accuracy of 0.9130 reveals the model's success on unseen data, while validation precision and recall values of 0.9061 and 0.9215 respectively showcase variations across epochs. These fluctuations may be attributed to learning rate adjustments or data variability during training. Table 4.5 represents the DenseNet121 results on the X-ray Images.

\begin{table}[htbp]
\centering
\caption{Performance Metrics of DenseNet121 on X-ray Images}
\begin{tabular}{|c|c|c|}
\hline
\textbf{Performance Metric} & \textbf{Training Result} & \textbf{Validation Result} \\
\hline
Loss & 0.1863 & 0.2211 \\
\hline
Accuracy & 0.9162 & 0.9130 \\
\hline
Precision & 0.9303 & 0.9061 \\
\hline
Recall & 0.8999 & 0.9215 \\
\hline
\end{tabular}
\end{table}

\subsection{DenseNet169}
\subsubsection{Results of DenseNet169 on CT Scan Images}
The DenseNet169 pre-trained model was utilized to analyze CT scan images of COVID-19 cases, resulting in the following evaluation metrics. The training process produced a relatively low loss of 0.0072, indicating that the model effectively learned the patterns within the data. The accuracy achieved during training was 99.72 percent, signifying the proportion of correctly classified instances in the training. The precision, which measures the accuracy of positive predictions, reached 99.70 percent, implying that the model had a high rate of correctly identifying true positive cases. The recall, also known as the true positive rate, stood at 99.75 percent, demonstrating the model's ability to capture a significant portion of actual positive cases. Figure 4.5 represents the DenseNet169 results on the CT scan images.

\begin{figure}[ht]
    \centering
    \includegraphics[width=0.8\textwidth]{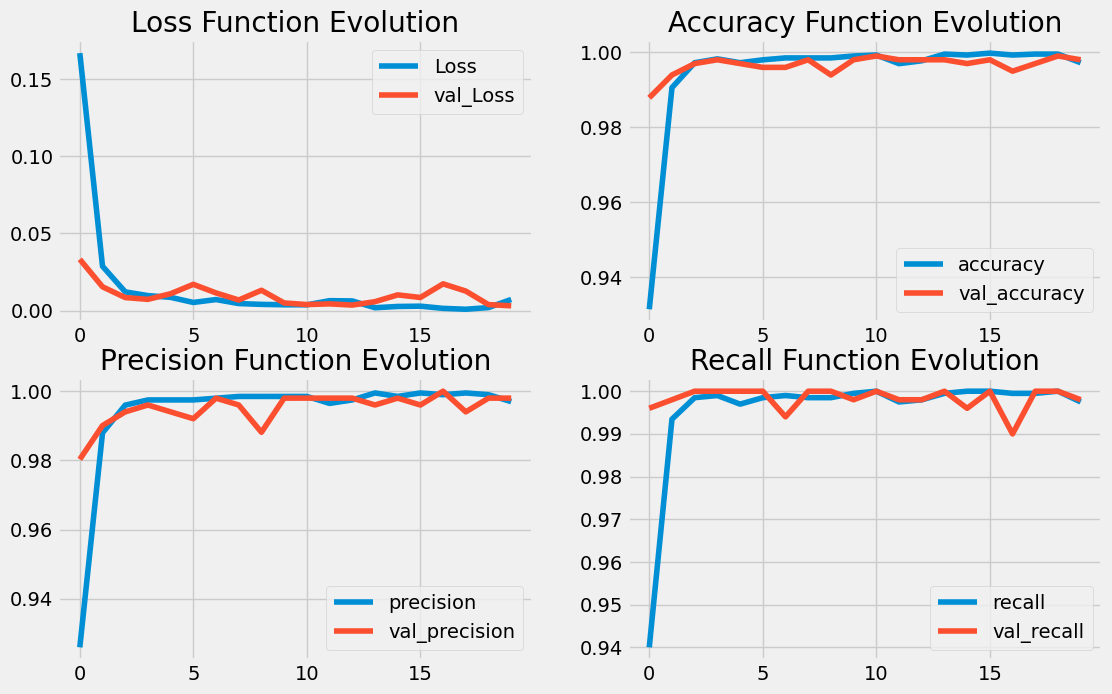}
    \caption{Results of DenseNet169 on CT scan Images}
\end{figure}

For validation, the model exhibited promising performance as well. The validation loss was 0.0031, showcasing the generalization capabilities of the model on unseen data. The validation accuracy of 99.80 percent demonstrated the model's ability to generalize its predictions to new, unseen cases. Both precision and recall were at 99.80 percent, emphasizing the model's capacity to maintain a high level of accuracy in predicting positive cases while minimizing false negatives. This implies that the model is efficient in capturing relevant information from the CT scan images, resulting in reliable predictions of COVID-19 cases. Table 4.6 represents the DenseNet169 results on the CT scan images.

\begin{table}[htbp]
\centering
\caption{Performance Metrics of DenseNet169 on CT Scan Images}
\begin{tabular}{|c|c|c|}
\hline
\textbf{Performance Metric} & \textbf{Training Result} & \textbf{Validation Result} \\
\hline
Loss & 0.0072 & 0.0031 \\
\hline
Accuracy & 0.9972 & 0.9980 \\
\hline
Precision & 0.9970 & 0.9980 \\
\hline
Recall & 0.9975 & 0.9980 \\
\hline
\end{tabular}
\end{table}

\subsubsection{Results of DenseNet169 on X-ray Images}

The employment of the DenseNet169 pretrained model for COVID-19 X-ray image analysis resulted in insightful outcomes. While training, the model exhibited a commendable accuracy of 94.48 percent. Notably, precision and recall, which gauge the model's aptitude in accurately identifying positive cases and the proportion of actual positives detected, displayed fluctuating patterns across epochs mspecifically in the validation dataset. This variability primarily arose during validation, implying potential challenges in maintaining consistent precision and recall during model evaluation. Importantly, these fluctuations in precision and recall were distinct from the testing, rather than the training. Figure 4.6 represents the DenseNet169 results on the X-ray Images.

\begin{figure}[ht]
    \centering
    \includegraphics[width=0.8\textwidth]{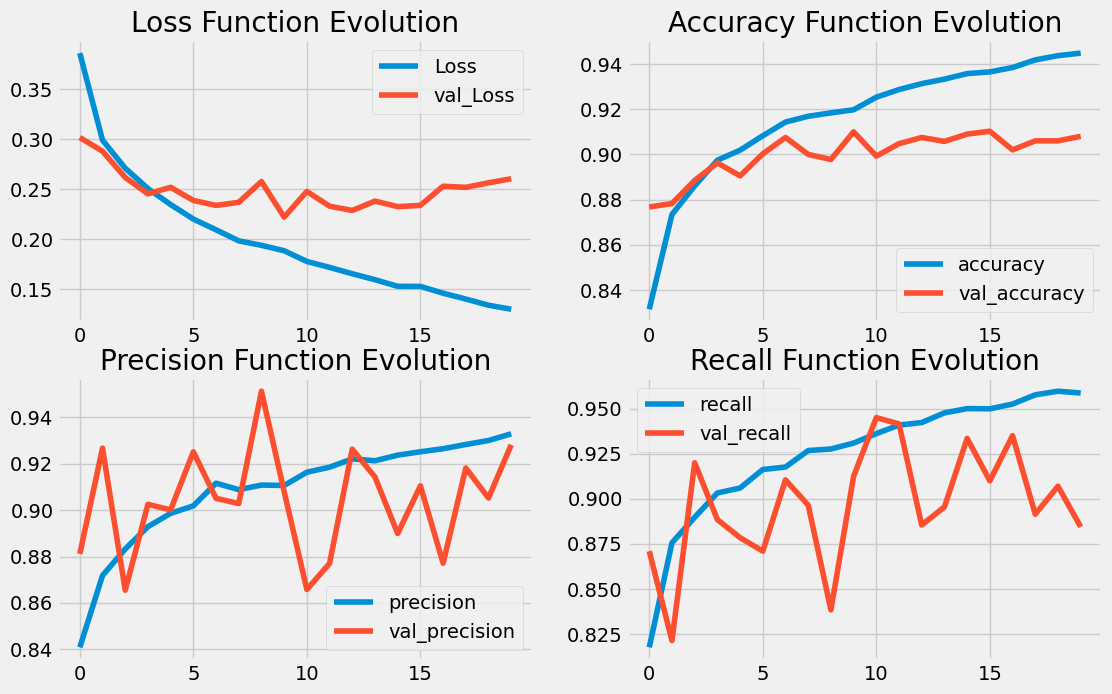}
    \caption{Results of DenseNet169 on X-ray images}
\end{figure}

Furthermore, it is worth noting that the loss metric showed an intriguing trend. Following the 12th epoch, the loss value started to increase gradually. This observation might point to the emergence of overfitting, a phenomenon wherein the model begins to tailor itself excessively to the training data and thus struggles to generalize well to new, unseen data.

During the validation phase, the model's performance was relatively diminished, yielding an accuracy of 90.80 percent. The variations in precision and recall values observed between epochs in the validation set highlight the model's sensitivity to certain features or instances within the validation. Particularly intriguing is the proximity of precision and recall values, indicating a balanced trade-off between minimizing false positives and false negatives. Table 4.7 represents the DenseNet169 results on the X-ray Images.

\begin{table}[htbp]
\centering
\caption{Performance Metrics of DenseNet169 on X-ray Images}
\begin{tabular}{|c|c|c|}
\hline
\textbf{Performance Metric} & \textbf{Training Result} & \textbf{Validation Result} \\
\hline
Loss & 0.1302 & 0.2606 \\
\hline
Accuracy & 0.9448 & 0.9080 \\
\hline
Precision & 0.9329 & 0.9281 \\
\hline
Recall & 0.9586 & 0.8845 \\
\hline
\end{tabular}
\end{table}

\subsection{DenseNet201}
\subsubsection{Results of DenseNet201 on CT Scan Images}
When utilized for CT scan images of COVID-19, the DenseNet201 pretrained model demonstrated encouraging outcomes. During training, the model achieved a low loss value of 0.0033 and a high accuracy of 99.95 percent. Moreover, the precision achieved was a perfect 100 percent, indicating that almost all positive predictions made by the model were accurate. The recall value of 99.90 percent suggests that the model effectively identified the majority of true positive cases. Figure 4.7 represents the DenseNet201 results on the CT scan images.

\begin{figure}[ht]
    \centering
    \includegraphics[width=0.8\textwidth]{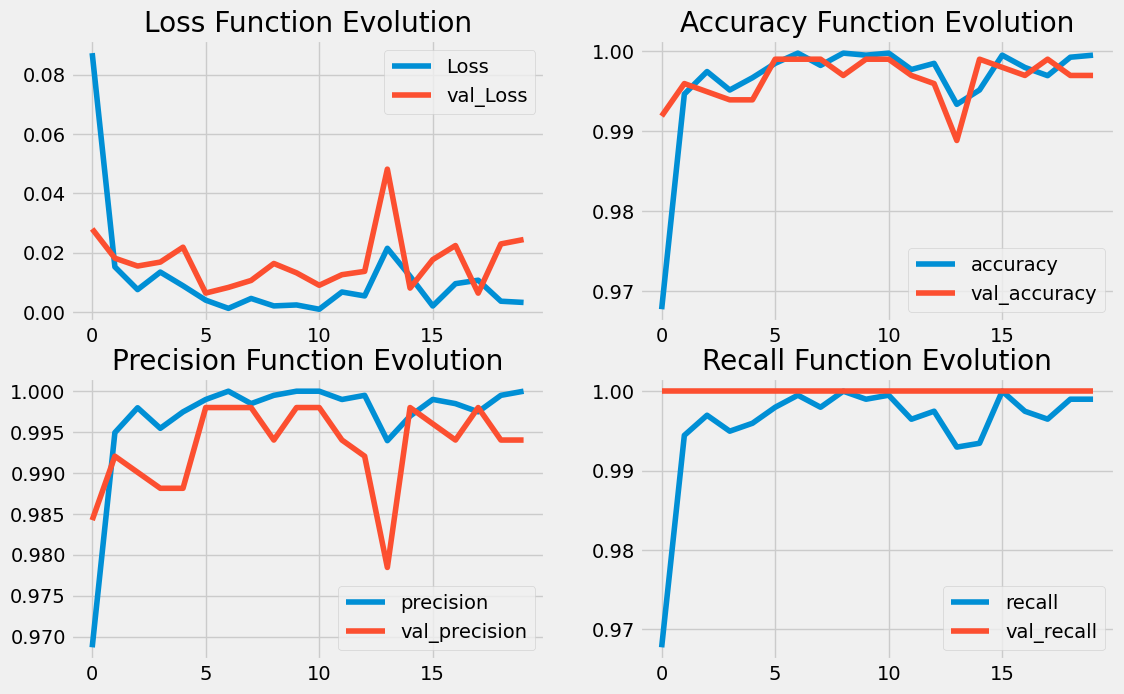}
    \caption{Results of DenseNet201 on CT scan Images}
\end{figure}

Nevertheless, during validation, the model demonstrated consistently high performance. The validation loss marginally increased to 0.0245, while the accuracy remained impressive at 99.70 percent. The validation precision of 99.40 percent signifies a high proportion of accurate positive predictions, and the validation recall of 100 percent affirms the model's ability to correctly identify all actual positive cases. Table 4.8 represents the DenseNet201 results on the CT scan images.

\begin{table}[htbp]
\centering
\caption{Performance Metrics of DenseNet201 on CT Scan Images}
\begin{tabular}{|c|c|c|}
\hline
\textbf{Performance Metric} & \textbf{Training Result} & \textbf{Validation Result} \\
\hline
Loss & 0.0033 & 0.0245 \\
\hline
Accuracy & 0.9995 & 0.9970 \\
\hline
Precision & 1.0000 & 0.9940 \\
\hline
Recall & 0.9990 & 1.0000 \\
\hline
\end{tabular}
\end{table}

\subsubsection{Results of DenseNet201 on X-ray Images}
Implementing the DenseNet201 pre-trained model on COVID-19 X-ray images led to notable outcomes. The training phase showcased a commendable accuracy of 90.69 percent, accompanied by a relatively moderate loss value of 0.2507. The precision during training, measured at 93.76 percent, highlights the model's proficiency in accurately identifying positive cases among predictions. The corresponding recall rate of 87.19 percent signifies the model's ability to successfully capture actual positive instances.

\begin{figure}[ht]
    \centering
    \includegraphics[width=0.8\textwidth]{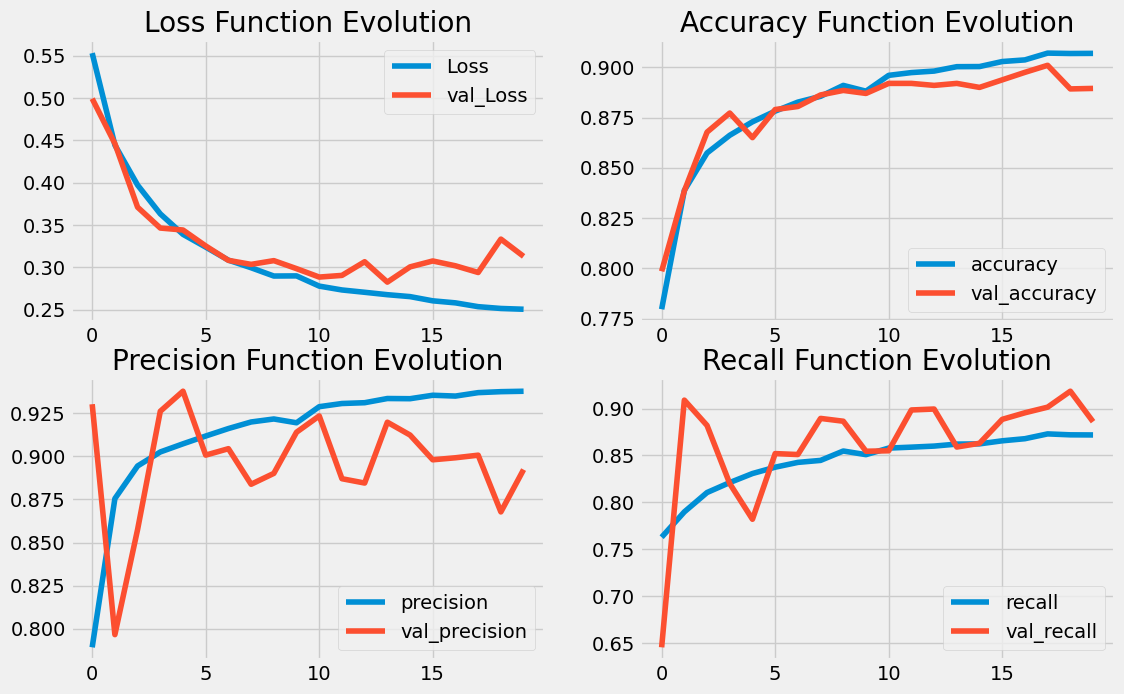}
    \caption{Results of DenseNet201 on X-ray Images}
\end{figure}

A notable observation pertains to the validation phase, where the precision metric exhibited a significant drop and subsequent recovery. Specifically, the validation precision experienced a noticeable decline, reaching its lowest point around the 12th epoch. Figure 4.8 represents the DenseNet201 results on the X-ray images.\newpage
\newpage
However, the model's performance rebounded remarkably, achieving a validation precision of 89.22 percent by the conclusion of the training process. Alongside this trend, the validation recall remained consistently strong throughout, measuring at 88.60 percent. These results collectively underscore the model's robustness in identifying positive cases, while the variation in validation precision showcases the model's adaptive learning over epochs. Table 4.9 represents the DenseNet201 results on the X-ray images.
\begin{table}[htbp]
\centering
\caption{Performance Metrics of DenseNet201 on X-ray Images}
\begin{tabular}{|c|c|c|}
\hline
\textbf{Performance Metric} & \textbf{Training Result} & \textbf{Validation Result} \\
\hline
Loss & 0.2507 & 0.3131 \\
\hline
Accuracy & 0.9069 & 0.8895 \\
\hline
Precision & 0.9376 & 0.8922 \\
\hline
Recall & 0.8719 & 0.8860 \\
\hline
\end{tabular}
\end{table}
\newpage

\subsection{VGG16}
\subsubsection{Results of VGG16 on CT Scan Images}
The utilization of the VGG16 pretrained model on COVID-19 CT images yielded highly favorable outcomes. The model achieved perfect accuracy, precision, and recall during training, indicating its ability to effectively differentiate between COVID-19 cases and non-COVID cases in the CT images. Figure 4.9 represents the VGG16 results on the CT scan images.

\begin{figure}[ht]
    \centering
    \includegraphics[width=0.8\textwidth]{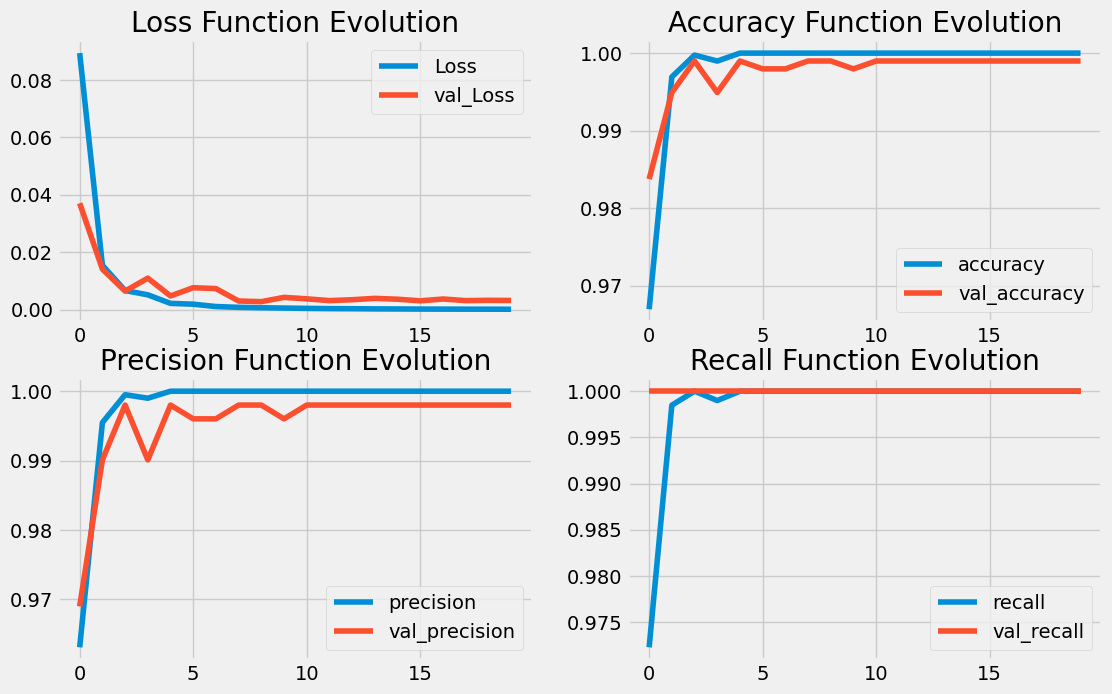}
    \caption{Results of VGG16 on CT Scan Images}
\end{figure}

During validation, the model maintained excellent performance, with only a slight increase in loss and a small drop in accuracy, both of which remain impressive. The validation precision of 99.80 percent demonstrates a high proportion of accurate positive predictions, and the validation recall of 100 percent indicates the model's proficiency in identifying all actual positive cases. Table 4.10 represents the VGG16 results on the CT scan images.

\begin{table}[htbp]
\centering
\caption{Performance Metrics of VGG16 on CT Scan Images}
\begin{tabular}{|c|c|c|}
\hline
\textbf{Performance Metric} & \textbf{Training Result} & \textbf{Validation Result} \\
\hline
Loss & 0.0001 & 0.0032 \\
\hline
Accuracy & 1.0000 & 0.9990 \\
\hline
Precision & 1.0000 & 0.9980 \\
\hline
Recall & 1.0000 & 1.0000 \\
\hline
\end{tabular}
\end{table}

\subsubsection{Results of VGG16 on X-ray Images}
The application of the VGG16 pre-trained model on X-ray images representing COVID-19 cases led to moderate outcomes. The recorded loss of 0.0818 and an accuracy of 0.9694 indicate successful learning and prediction during the training phase. However, during the validation phase, an interesting phenomenon occurred after 10 epochs. Both the validation accuracy and precision experienced a decline, while the validation loss increased. It's important to emphasize that this effect was specifically observed in the validation results, indicating potential fluctuations in the model's generalization to unseen data. Figure 4.10 represents the VGG16 results on the X-ray images.

\begin{figure}[ht]
    \centering
    \includegraphics[width=0.8\textwidth]{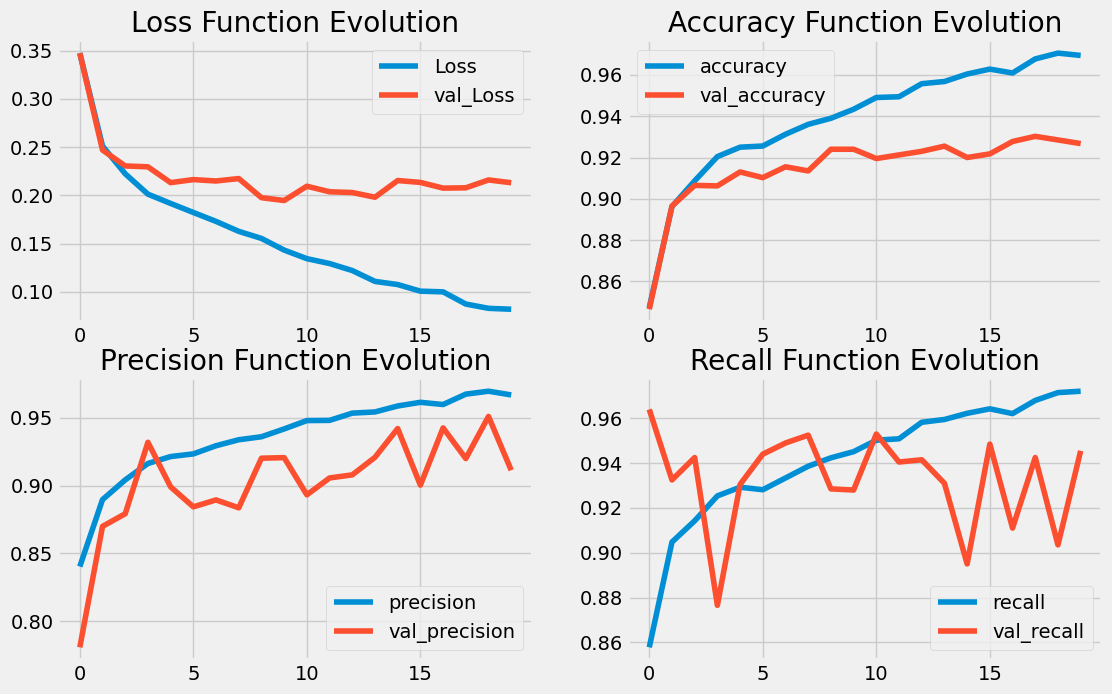}
    \caption{Results of VGG16 on X-ray Images}
\end{figure}

Contrastingly, the training process continued smoothly and showcased consistent progress, reaffirming the model's capacity to effectively learn from the training dataset. The training process extended for a total of 20 epochs, reflecting the model's dedication to refining its predictive capabilities over a substantial number of training iterations. This instance underscores the dynamic nature of model training and the intricate interplay between various evaluation metrics during different phases of the training process. Table 4.11 represents the VGG16 results on the X-ray images.

\begin{table}[htbp]
\centering
\caption{Performance Metrics of VGG16 on X-ray Images}
\begin{tabular}{|c|c|c|}
\hline
\textbf{Performance Metric} & \textbf{Training Result} & \textbf{Validation Result} \\
\hline
Loss & 0.0818 & 0.2131 \\
\hline
Accuracy & 0.9694 & 0.9268 \\
\hline
Precision & 0.9669 & 0.9113 \\
\hline
Recall & 0.9721  & 0.9455 \\
\hline
\end{tabular}
\end{table}

\subsection{VGG19}
\subsubsection{Results of VGG19 on CT Scan Images}
VGG19 pre-trained model to CT images of COVID-19 cases led to compelling outcomes. During training, the model achieved a minimal loss of 0.0005, indicating its ability to make accurate predictions. The accuracy, precision, and recall rates were all perfect, with values of 1.0000, showcasing the model's ability to correctly classify instances. Figure 4.11 represents the VGG19 results on the CT scan images.\newpage

\begin{figure}[ht]
    \centering
    \includegraphics[width=0.7\textwidth]{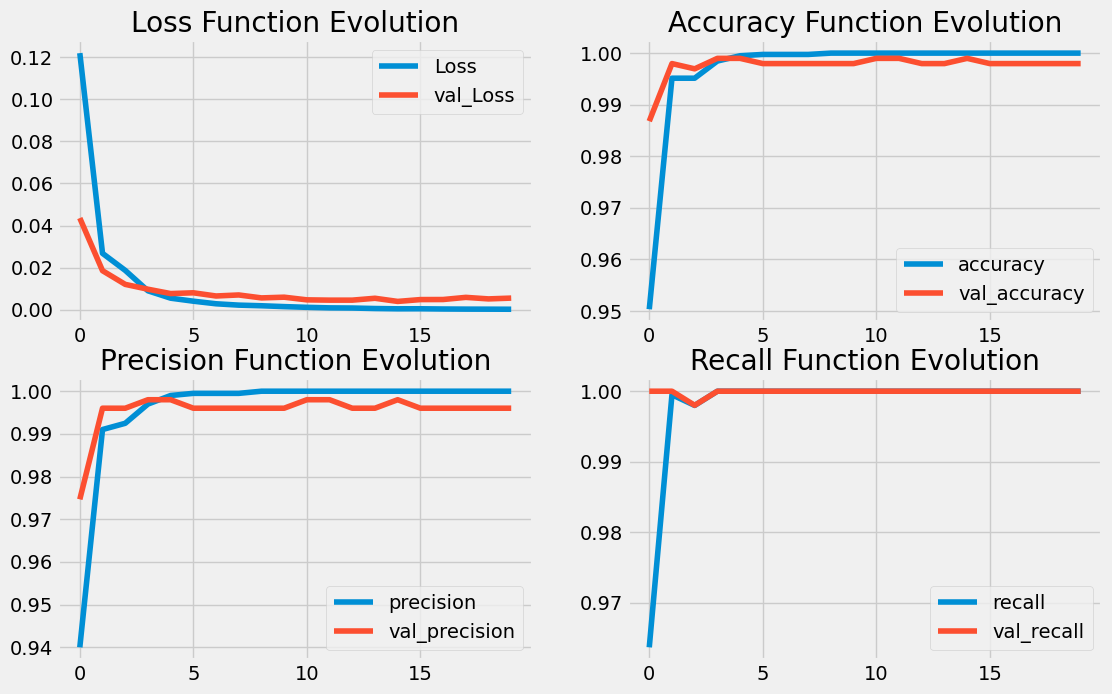}
    \caption{Results of VGG19 on CT scan Images}
\end{figure}

In the validation phase, the model maintained its high performance, with a validation loss of 0.0055 and a validation accuracy of 0.9980. The validation precision and recall rates were also commendable, with values of 0.9960 and 1.0000, respectively. This consistent performance across training and validation demonstrates the robustness of the VGG19 model in accurately identifying COVID-19 cases from CT images. Table 4.12 represents the VGG19 results on the CT scan images.

\begin{table}[htbp]
\centering
\caption{Performance Metrics of VGG19 on CT Scan Images}
\begin{tabular}{|c|c|c|}
\hline
\textbf{Performance Metric} & \textbf{Training Result} & \textbf{Validation Result} \\
\hline
Loss & 0.0005 & 0.0055 \\
\hline
Accuracy & 1.0000 & 0.9980 \\
\hline
Precision & 1.0000 & 0.9960 \\
\hline
Recall & 1.0000 & 1.0000 \\
\hline
\end{tabular}
\end{table}

\subsubsection{Results of VGG19 on X-ray Images}
The utilization of the VGG19 pre-trained model on the X-ray images for COVID-19 classification. The achieved loss of 0.1791 and accuracy of 0.9274 on the training data indicate the model's proficiency in making accurate predictions. The precision value of 0.9232 signifies the proportion of correctly predicted positive instances among all predicted positives, while the recall value of 0.9325 highlights the model's ability to identify actual positive cases. Figure 4.12 represents the VGG19 results on the X-ray images. \newpage

\begin{figure}[ht]
    \centering
    \includegraphics[width=0.8\textwidth]{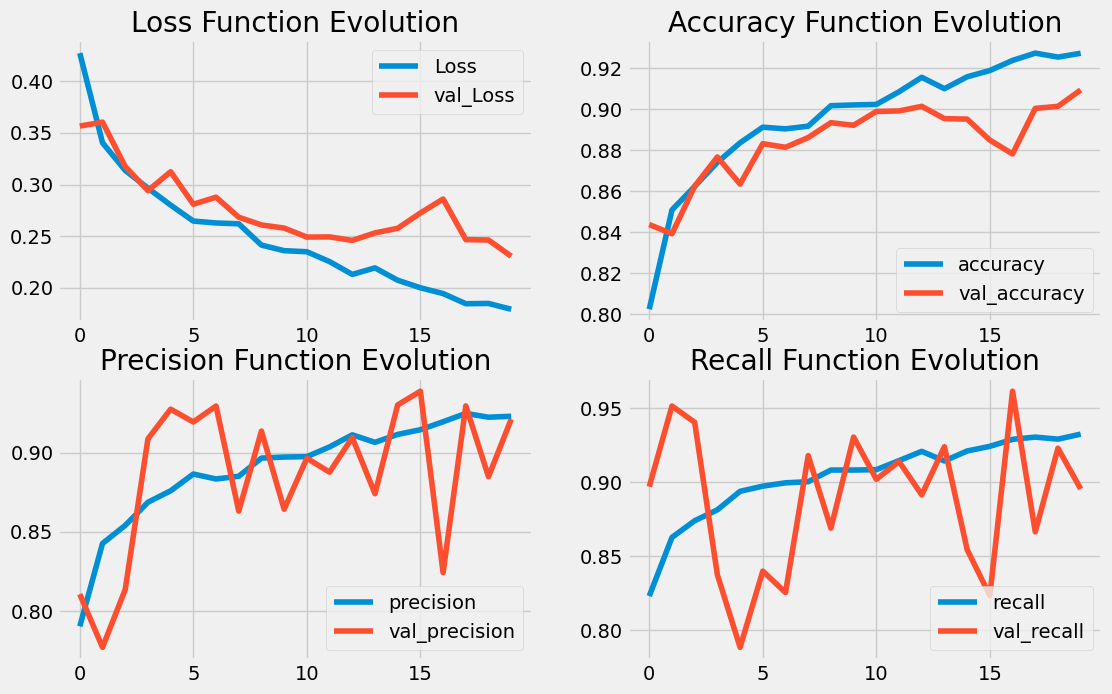}
    \caption{Results of VGG19 on X-ray imagest}
\end{figure}

During validation, the model exhibited a relatively stable performance, as indicated by the validation loss of 0.2305 and validation accuracy of 0.9095. However, a subtle change in the validation results was observed after 15 epochs, particularly in the loss and accuracy metrics. Notably, the validation precision and validation recall values varied on the testing data, suggesting some fluctuations in the model's prediction capabilities for positive instances. It's worth mentioning that the training process showed consistent and satisfactory progress, indicating that the model was able to learn and generalize from the training effectively. Table 4.13 represents the VGG19 results on the X-ray images.

\begin{table}[htbp]
\centering
\caption{Performance Metrics of VGG19 on X-ray Images}
\begin{tabular}{|c|c|c|}
\hline
\textbf{Performance Metric} & \textbf{Training Result} & \textbf{Validation Result} \\
\hline
Loss & 0.1791 & 0.2305 \\
\hline
Accuracy & 0.9274 & 0.9095 \\
\hline
Precision & 0.9232 & 0.9213 \\
\hline
Recall & 0.9325  & 0.8955 \\
\hline
\end{tabular}
\end{table}

\subsection{Inception V3}
\subsubsection{Results of Inception V3 on CT Scan Images}
The utilization of the Inception V3 pre-trained model on CT images portraying COVID-19 cases yielded remarkable outcomes. During the training phase, the model achieved a flawless accuracy, precision, and recall, effectively identifying and classifying all instances of COVID-19. Figure 4.13 represents the inception V3 results on the CT scan images.
\newpage

\begin{figure}[ht]
    \centering
    \includegraphics[width=0.8\textwidth]{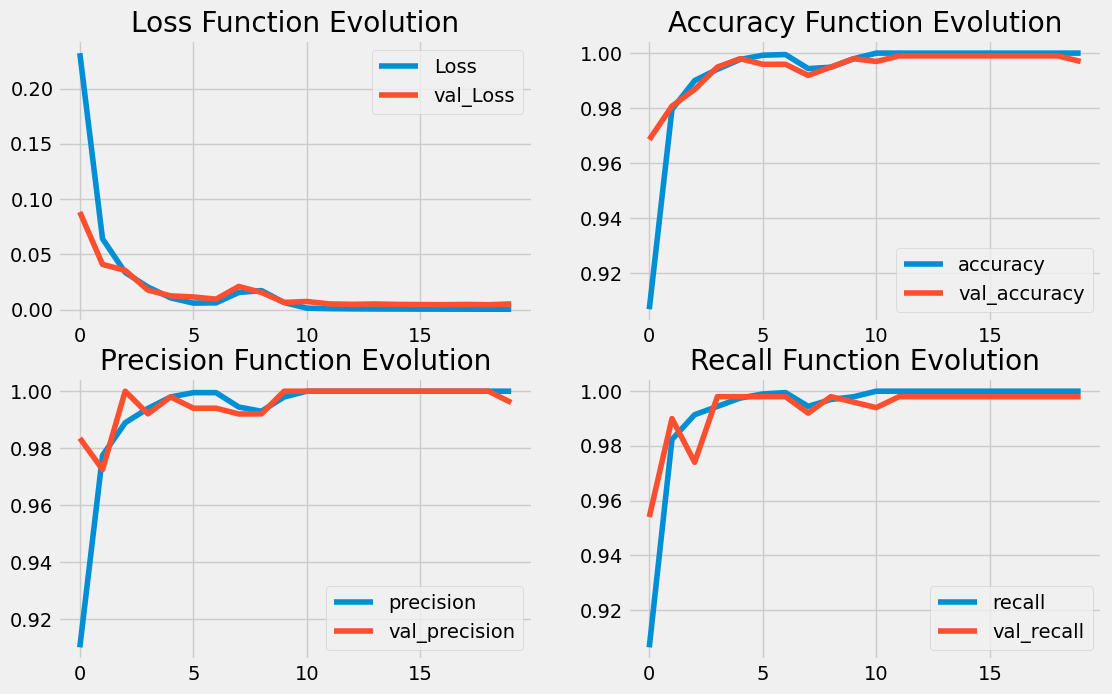}
    \caption{Results of Inception V3 on CT Scan Images}
\end{figure}

Notably, during the validation phase, the model's performance remained impressive, with an accuracy of 99.70 percent. Additionally, the precision score of 99.60 percent signifies the high accuracy of positive predictions, while the recall score of 99.80 percent indicates the model's proficiency in identifying actual positive cases. This consistent and robust performance across both training and validation phases underscores the model's efficacy in accurately diagnosing COVID-19 cases from CT images. Table 4.14 represents the inception V3 results on the CT scan images.

\begin{table}[htbp]
\centering
\caption{Performance Metrics of Inception V3 on CT Scan Images}
\begin{tabular}{|c|c|c|}
\hline
\textbf{Performance Metric} & \textbf{Training Result} & \textbf{Validation Result} \\
\hline
Loss & 0.0003 & 0.0051 \\
\hline
Accuracy & 1.0000 & 0.9970 \\
\hline
Precision & 1.0000 & 0.9960 \\
\hline
Recall & 1.0000 & 0.9980 \\
\hline
\end{tabular}
\end{table}

\subsubsection{Results of Inception V3 on X-ray Images}
The application of the Inception V3 pre-trained model on X-ray images of COVID-19 cases yielded results that were not as favorable as anticipated. The obtained accuracy, precision, and recall values were 0.8075, 0.7981, and 0.8232, respectively, for the training set. Figure 4.14 represents the inception V3 results on the X-ray images.

\newpage

\begin{figure}[ht]
    \centering
    \includegraphics[width=0.8\textwidth]{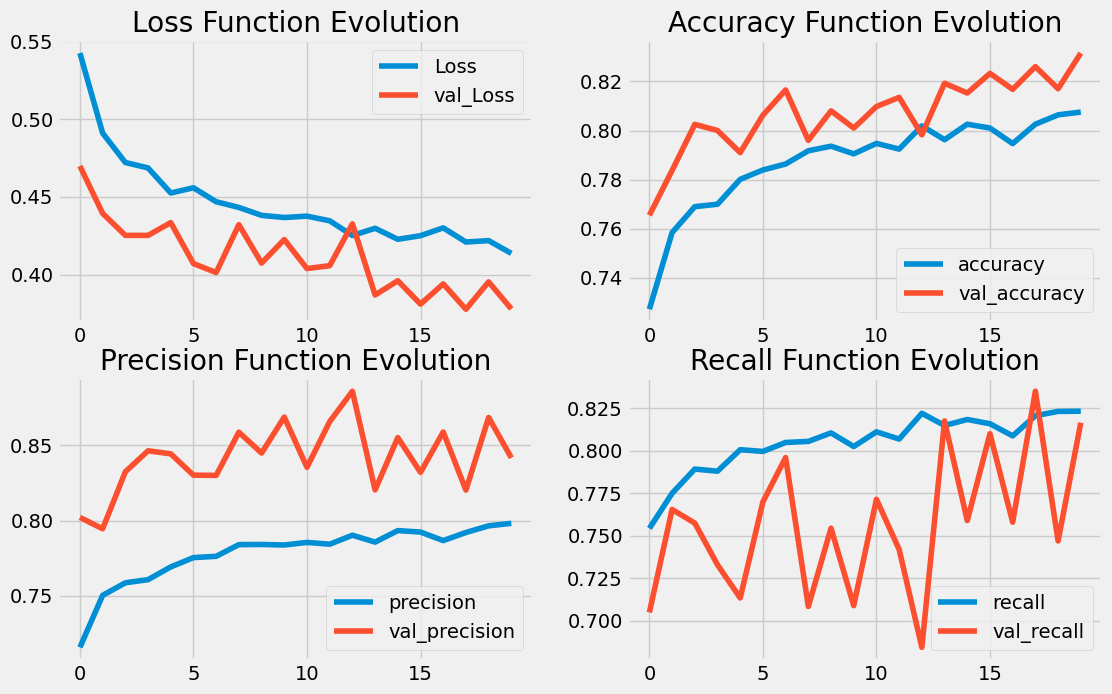}
    \caption{Results of Inception V3 on X-ray Images}
\end{figure}

In the validation phase, the model exhibited a loss of 0.3781 and achieved an accuracy of 0.8315, a precision of 0.8418, and a recall of 0.8165. Despite various attempts to enhance the model's performance through hyperparameter tuning, including regularization, dropout layers, batch normalization, and data augmentation, the improvement remained limited. An interesting observation is that the validation metrics (accuracy and precision) were consistently higher than their training counterparts. Table 4.15 represents the inception V3 results on the X-ray images.

\begin{table}[htbp]
\centering
\caption{Performance Metrics of Inception V3 on X-ray Images}
\begin{tabular}{|c|c|c|}
\hline
\textbf{Performance Metric} & \textbf{Training Result} & \textbf{Validation Result} \\
\hline
Loss & 0.4138 & 0.3781 \\
\hline
Accuracy & 0.8075 & 0.8315 \\
\hline
Precision & 0.7981 & 0.8418 \\
\hline
Recall & 0.8232 & 0.8165 \\
\hline
\end{tabular}
\end{table}

\subsection{Inception V4}
\subsubsection{Results of Inception V4 on CT Scan Images}
The implementation of the Inception V4 model from scratch on CT scan images depicting COVID-19 cases resulted in exceptional outcomes. During training, the model achieved a remarkable accuracy of 100 percent, indicating that it accurately classified all instances in the training set. The precision and recall scores of 1.0000 further confirm the model's ability to make correct positive predictions and capture all actual positive cases. Figure 4.15 represents the inception V4 results on the CT scan images.
\newpage

\begin{figure}[ht]
    \centering
    \includegraphics[width=0.8\textwidth]{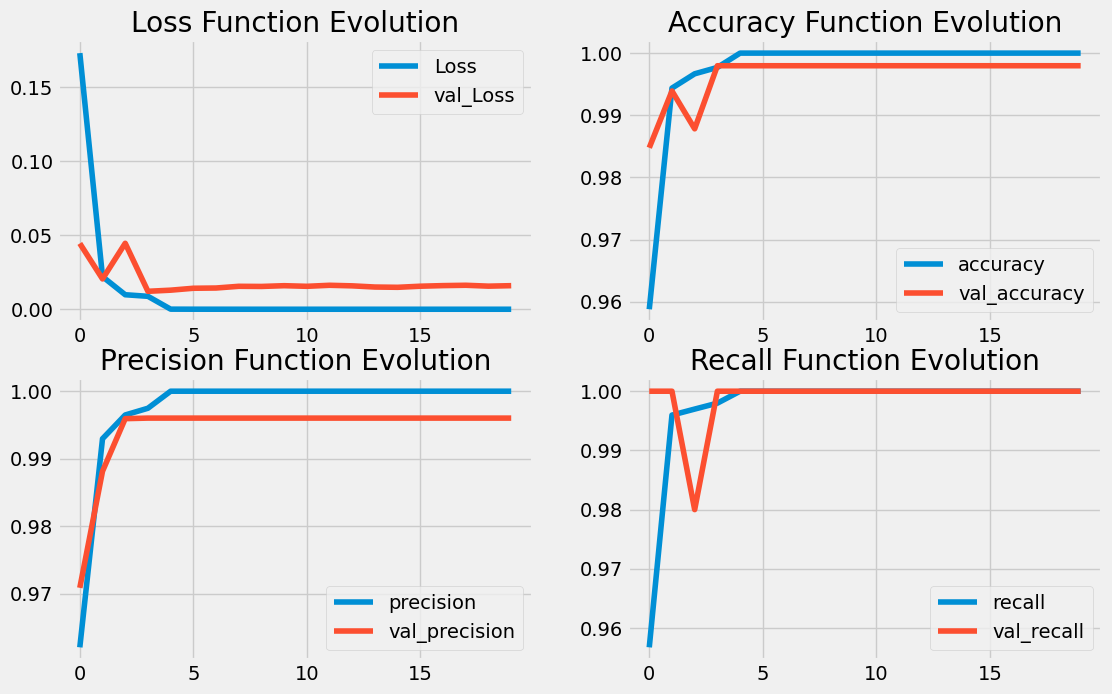}
    \caption{Results of Inception V4 on CT scan Images}
\end{figure}

The validation results were also promising, with an accuracy of 99.80 percent and a precision of 99.60 percent, demonstrating the model's strong generalization performance on unseen data. The slight decrease in precision in the validation set suggests that some positive predictions might not be as accurate as in the training set. Table 4.16 represents the inception V4 results on the CT scan images.

\begin{table}[htbp]
\centering
\caption{Performance Metrics of Inception V4 on CT Scan Images}
\begin{tabular}{|c|c|c|}
\hline
\textbf{Performance Metric} & \textbf{Training Result} & \textbf{Validation Result} \\
\hline
Loss & 0.0000001 & 0.0159 \\
\hline
Accuracy & 1.0000 & 0.9980 \\
\hline
Precision & 1.0000 & 0.9960 \\
\hline
Recall & 1.0000 & 1.0000 \\
\hline
\end{tabular}
\end{table}

\subsubsection{Results of Inception V4 on X-ray Images}
Inception V4 model from scratch Implemented on the X-ray images. Despite the absence of a pre-trained model, the achieved accuracy of 0.9943 on the training data showcases the model's ability to make accurate predictions. The precision value of 0.9939 highlights the high proportion of correctly predicted positive instances among all predicted positives, while the recall value of 0.9948 underscores the model's capability to identify actual positive cases effectively. Figure 4.16 represents the inception V4 results on the X-ray images.

\newpage

\begin{figure}[ht]
    \centering
    \includegraphics[width=0.8\textwidth]{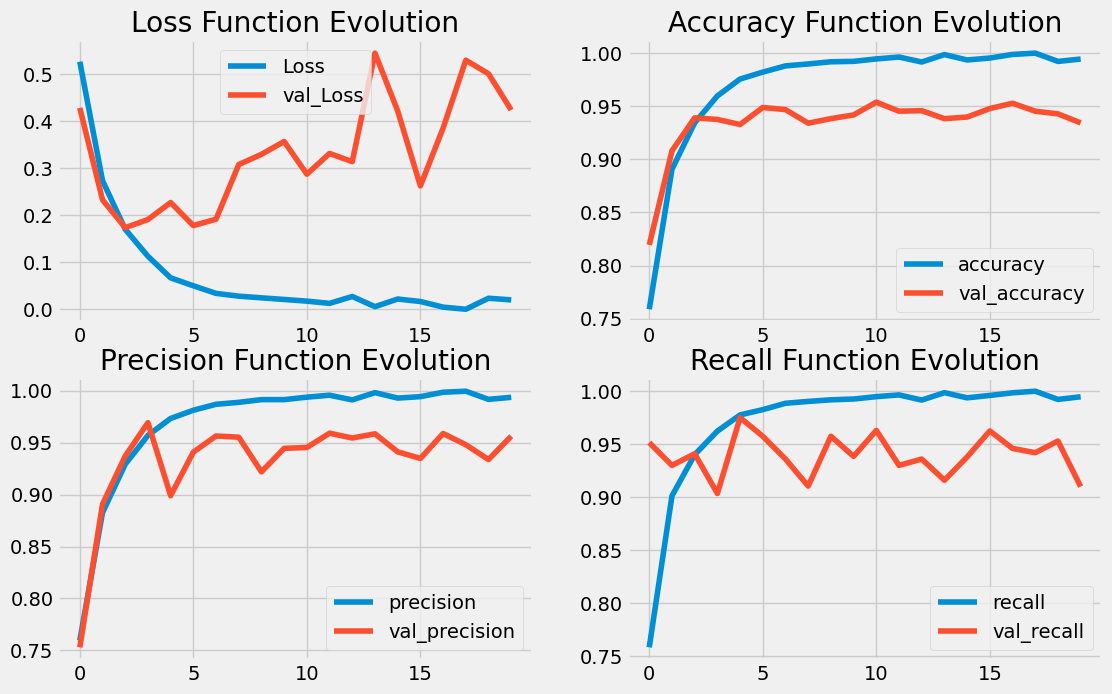}
    \caption{Results of Inception V4 on X-ray Images}
\end{figure}

During the validation process, the model exhibited certain fluctuations in performance. The validation loss of 0.4241 and validation accuracy of 0.9342 suggest that the model's performance on unseen data remained favorable but slightly less stable compared to the training data. Notably, after just three epochs, an increase in validation loss was observed, indicating a possible struggle of the model to generalize well on the validation. Despite this, the model managed to maintain a relatively high validation accuracy and precision, with values of 0.9564 and 0.9100 respectively, highlighting its capacity to identify positive instances accurately. Table 4.17 represents the inception V4 results on the X-ray images.

\begin{table}[htbp]
\centering
\caption{Performance Metrics of Inception V4 on X-ray Images}
\begin{tabular}{|c|c|c|}
\hline
\textbf{Performance Metric} & \textbf{Training Result} & \textbf{Validation Result} \\
\hline
Loss & 0.0203 & 0.4241 \\
\hline
Accuracy & 0.9943 & 0.9342 \\
\hline
Precision & 0.9939 & 0.9564 \\
\hline
Recall & 0.9948 & 0.9100 \\
\hline
\end{tabular}
\end{table}

\subsection{AlexNet}
\subsubsection{Results of AlexNet on CT Scan Images}
The implementation of AlexNet using Keras on CT scan images resulted in highly favorable outcomes. The model achieved perfect performance on both the training and validation. During training, the model exhibited a remarkably low loss of 0.000007, indicating accurate predictions. The accuracy, precision, and recall were all perfect with values of 1.0000, implying that all predictions were correct, and no positive cases were missed. Figure 4.17 represents the AlexNet results on the CT scan images.
\newpage

\begin{figure}[ht]
    \centering
    \includegraphics[width=0.7\textwidth]{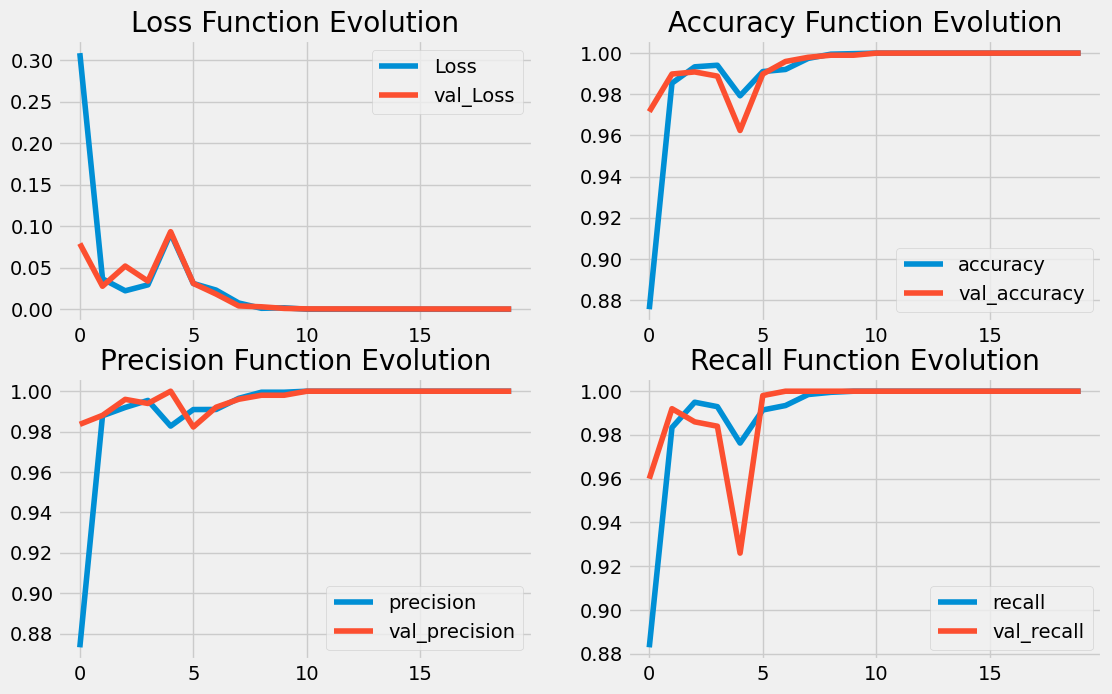}
    \caption{Results of AlexNet on CT scan Images}
\end{figure}

In the validation phase, the model's performance was equally impressive, achieving a very low loss of 0.0002 and perfect accuracy, precision, and recall values of 1.0000. These results indicate the model's capability to effectively differentiate between COVID-19 and non-COVID-19 cases in CT scan images. Table 4.18 represents the AlexNet results on the CT scan images.

\begin{table}[htbp]
\centering
\caption{Performance Metrics of AlexNet on CT Scan Images}
\begin{tabular}{|c|c|c|}
\hline
\textbf{Performance Metric} & \textbf{Training Result} & \textbf{Validation Result} \\
\hline
Loss & 0.0000007 & 0.0002 \\
\hline
Accuracy & 1.0000 & 1.000 \\
\hline
Precision & 1.0000 & 1.0000 \\
\hline
Recall & 1.0000 & 1.0000 \\
\hline
\end{tabular}

\end{table}

\subsubsection{Results of AlexNet on X-ray Images}
Implementing AlexNet using Keras on the X-ray images resulted in promising outcomes, albeit with some variations observed during the validation phase. During training, the model demonstrated effective learning with a relatively low loss of 0.0909, indicating accurate predictions. The accuracy, precision, and recall were also favorable, with values of 0.9651, 0.9601, and 0.9705, respectively. These values suggest the model's capability to distinguish COVID-19 cases from non-COVID-19 cases in X-ray images. Figure 4.18 represents the AlexNet results on the X-ray images.
\newpage

\begin{figure}[ht]
    \centering
    \includegraphics[width=0.8\textwidth]{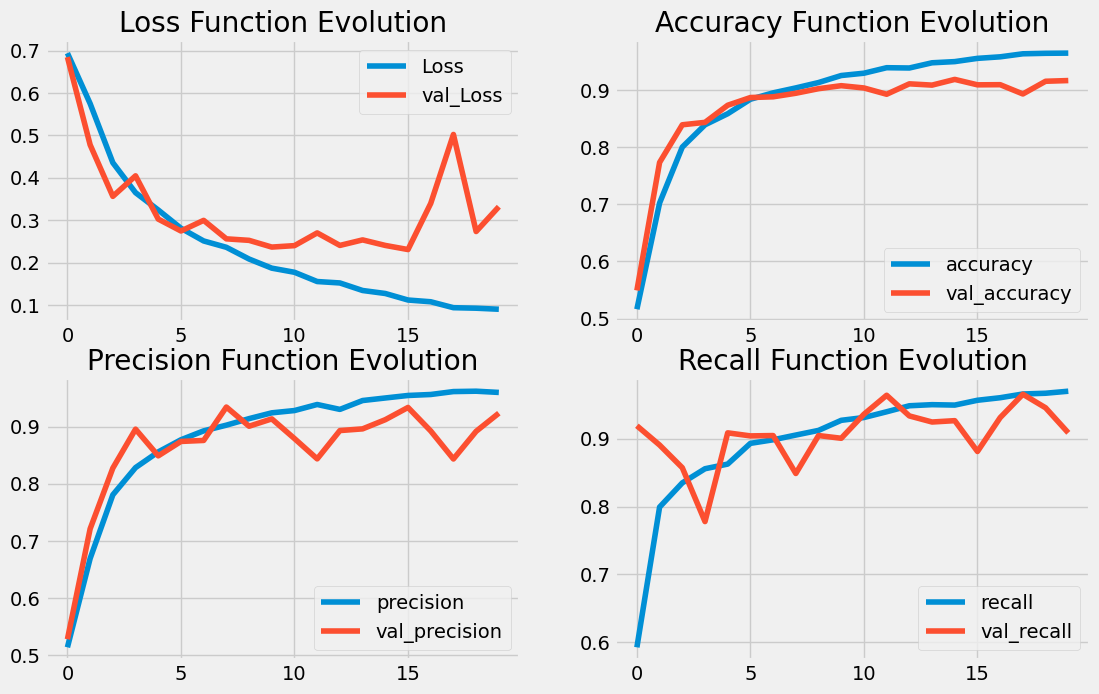}
    \caption{Results of AlexNet on X-ray Images}
\end{figure}

However, it's important to note that during the validation phase, a variation in the validation loss was observed after 10 epochs. This could be indicative of some instability in the model's generalization performance, possibly due to overfitting or other factors. Despite this variability, the validation loss remained within an acceptable range at 0.3323, and the validation accuracy, precision, and recall were also relatively high, with values of 0.9170, 0.9238, and 0.9090, respectively. This indicates that the model's performance was still favorable on the validation set, although some caution should be exercised due to the observed loss variation. Table 4.19 represents the AlexNet results on the X-ray images.

\begin{table}[htbp]
\centering
\caption{Performance Metrics of AlexNet on X-ray Images}
\begin{tabular}{|c|c|c|}
\hline
\textbf{Performance Metric} & \textbf{Training Result} & \textbf{Validation Result} \\
\hline
Loss & 0.0909 & 0.3323 \\
\hline
Accuracy & 0.9651 & 0.9170 \\
\hline
Precision & 0.9601 & 0.9238 \\
\hline
Recall & 0.9705 & 0.9090 \\
\hline
\end{tabular}
\end{table}

\subsection{Remarks and Conclusions}
This summary provides an overview of our study on the performance of various deep learning models for COVID-19 detection using medical imaging, specifically CT scan and X-ray images. The study evaluated a
range of pre-trained models, including CNN, DenseNet, VGG, Inception, and AlexNet, across diverse datasets. In this comparative study of deep learning models on the CT scan dataset, we not only assess their classification performance but also delve into execution time and memory utilization. The Convolutional Neural Network (CNN) model emerges as a top performer, boasting exceptional accuracy, precision, and recall, with minimal loss. Notably, the CNN executes swiftly and efficiently, making it an attractive choice for real-time applications.

Among the other models, DenseNet121, DenseNet169, and DenseNet201 exhibit commendable accuracy and generalization, while consuming reasonable computational resources. However, when considering VGG16,
VGG19, Inception V3, Inception V4, and AlexNet, we observe a trade-off. These models, known for their deep architectures and high-resolution image requirements, demand significantly more memory and execution time. While they achieve competitive results, their computational cost may be a limiting factor in resource-constrained environments

It’s worth highlighting that the VGG and AlexNet models, due to their depth and image resolution prerequisites, may not be the most practical choices when execution time and memory efficiency are critical considerations. Thus, the choice of model should align with specific application requirements, taking into account
not only classification performance but also computational constraints. Table 4.20 presents a comprehensive
comparison of various models based on their performance using CT scan Images. 

\begin{table}[ht]
\centering
\caption{Performance Metrics for CT Scan Images}
\begin{tabular}{|c|c|c|c|}
\hline
\textbf{Models} & \textbf{Metrics} & \textbf{Training Results} & \textbf{Testing Results} \\
\hline
\hline
CNN & Accuracy & 1.0000 & 0.9989 \\
    & Precision & 1.0000 & 0.9980 \\
    & Recall & 1.0000 & 1.0000 \\
    & Loss & 0.0017 & 0.0037 \\
\hline
DenseNet121 & Accuracy & 0.9974 & 0.9990 \\
            & Precision & 0.9965 & 0.9980 \\
            & Recall & 0.9985 & 1.0000 \\
            & Loss & 0.0080 & 0.0041 \\
\hline
DenseNet169 & Accuracy & 0.9972 & 0.9980 \\
            & Precision & 0.9970 & 0.9980 \\
            & Recall & 0.9975 & 0.9980 \\
            & Loss & 0.0072 & 0.0031 \\
\hline
DenseNet201 & Accuracy & 0.9995 & 0.9970 \\
            & Precision & 1.0000 & 0.9940 \\
            & Recall & 0.9990 & 1.0000 \\
            & Loss & 0.0033 & 0.0245 \\
\hline
VGG16 & Accuracy & 1.0000 & 0.9990 \\
      & Precision & 1.0000 & 0.9980 \\
      & Recall & 1.0000 & 1.0000 \\
      & Loss & 0.0001 & 0.0032 \\
\hline
VGG19 & Accuracy & 1.0000 & 0.9980 \\
      & Precision & 1.0000 & 0.9960 \\
      & Recall & 1.0000 & 1.0000 \\
      & Loss & 0.0005 & 0.0055 \\
\hline
Inception V3 & Accuracy & 1.0000 & 0.9970 \\
             & Precision & 1.0000 & 0.9960 \\
             & Recall & 1.0000 & 0.9980 \\
             & Loss & 0.0003 & 0.0051 \\
\hline
Inception V4 & Accuracy & 1.0000 & 0.9980 \\
             & Precision & 1.0000 & 0.9960 \\
             & Recall & 1.0000 & 1.0000 \\
             & Loss & 0.0000 & 0.0159 \\
\hline
AlexNet & Accuracy & 1.0000 & 1.0000 \\
        & Precision & 1.0000 & 1.0000 \\
        & Recall & 1.0000 & 1.0000 \\
        & Loss & 0.0000 & 0.0002 \\
\hline
\end{tabular}
\end{table}
\newpage

In the context of the X-ray image, the performance of various deep learning models provides valuable insights into their effectiveness for medical image classification. Table 4.21 presents a comprehensive comparison of various models based on their performance using X-ray data.

\begin{table}[ht]
\centering
\caption{Performance Metrics for X-ray Images}
\begin{tabular}{|c|c|c|c|}
\hline
\textbf{Model} & \textbf{Metric} & \textbf{Training Result} & \textbf{Validation Result} \\
\hline
CNN & Accuracy & 0.9950 & 0.9722 \\
    & Precision & 0.9951 & 0.9753 \\
    & Recall & 0.9949 & 0.9690 \\
    & Loss & 0.0154 & 0.1171 \\
\hline
DenseNet121 & Accuracy & 0.9162 & 0.9130 \\
            & Precision & 0.9303 & 0.9061 \\
            & Recall & 0.8999 & 0.9215 \\
            & Loss & 0.1863 & 0.2211 \\
\hline
DenseNet169 & Accuracy & 0.9448 & 0.9080 \\
            & Precision & 0.9329 & 0.9281 \\
            & Recall & 0.9586 & 0.8845 \\
            & Loss & 0.1302 & 0.2606 \\
\hline
DenseNet201 & Accuracy & 0.9069 & 0.8895 \\
            & Precision & 0.9376 & 0.8922 \\
            & Recall & 0.8719 & 0.8860 \\
            & Loss & 0.2507 & 0.3131 \\
\hline
VGG16 & Accuracy & 0.9694 & 0.9268 \\
      & Precision & 0.9669 & 0.9113 \\
      & Recall & 0.9721 & 0.9455 \\
      & Loss & 0.0818 & 0.2131 \\
\hline
VGG19 & Accuracy & 0.9274 & 0.9095 \\
      & Precision & 0.9232 & 0.9213 \\
      & Recall & 0.9325 & 0.8955 \\
      & Loss & 0.1791 & 0.2305 \\
\hline
Inception V3 & Accuracy & 0.8075 & 0.8315 \\
             & Precision & 0.7981 & 0.8418 \\
             & Recall & 0.8232 & 0.8165 \\
             & Loss & 0.4138 & 0.3781 \\
\hline
Inception V4 & Accuracy & 0.9943 & 0.9342 \\
             & Precision & 0.9939 & 0.9564 \\
             & Recall & 0.9948 & 0.9100 \\
             & Loss & 0.0203 & 0.4241 \\
\hline
AlexNet & Accuracy & 0.9651 & 0.9170 \\
        & Precision & 0.9601 & 0.9238 \\
        & Recall & 0.9705 & 0.9090 \\
        & Loss & 0.0909 & 0.3323 \\
\hline
\end{tabular}
\end{table}

In terms of accuracy, the CNN, DenseNet121, and Inception V4 models exhibit the highest levels during both training and validation, surpassing 90 percent. This signifies their ability to correctly classify X-ray images. While other models like DenseNet169 and DenseNet201 also achieve impressive accuracy, they tend to hover just below the 90 percent mark, suggesting a slightly lower overall performance.

Precision, which measures the accuracy of positive predictions, demonstrates a similar pattern. CNN, DenseNet121, and Inception V4 consistently achieve precision scores above 90 percent, indicating their proficiency in minimizing false positives. Meanwhile, DenseNet169 and DenseNet201 maintain respectable precision levels but exhibit slightly more variably

When it comes to recall, the ability to identify all relevant instances, CNN, DenseNet121, and InceptionV4 once again stand out, with recall values consistently exceeding 90 percent. This implies that these models excel in correctly identifying relevant findings within X-ray images. DenseNet169 and DenseNet201, although strong performers, tend to have slightly lower recall values.

Loss, representing the convergence of the training process, aligns with these trends. Models with higher accuracy, precision, and recall tend to have lower loss values, indicating more effective convergence during training.

It’s important to note that the choice of the most suitable model depends on the specific medical imaging task and computational resources available. Models like Inception V4 and AlexNet, which demand more computational power due to the dataset’s higher resolution, might be preferred when the highest level of accuracy and recall is critical. In contrast, models like DenseNet121 and DenseNet169 offer a good balance of performance and computational efficiency.

In conclusion, while all models show promising results for X-ray image classification, CNN, DenseNet121, and Inception V4 shine with consistently high accuracy, precision, and recall scores. DenseNet169 and DenseNet201 are also strong contenders, offering a slightly different trade-off between accuracy and computational demands.
The choice ultimately hinges on the specific requirements and resources available for the medical image classification task at hand.

Our analysis yielded the following key findings:
\begin{itemize}
    \item \textbf{Model Diversity}: Different models demonstrated varying levels of success in COVID-19 detection, emphasizing the importance of tailoring model selection to the specific dataset and imaging modality.
    
    \item \textbf{Performance on CT Scans}: Models generally excelled in performance when applied to CT scan images. CNN, AlexNet, and Inception models achieved high accuracy, precision, and recall, with minimal overfitting.
    
    \item \textbf{Performance on X-ray Images}: Model performance exhibited more variability when dealing with X-ray images. While many models produced favorable results, some displayed signs of overfitting, warranting caution.
    
    \item \textbf{Generalization Matters}: The ability of models to generalize to unseen data emerged as a critical factor. Models that maintained strong performance on validation data were considered more reliable for real-world deployment.
    
    \item \textbf{Clinical Validation}: While high model accuracy is valuable, it should be complemented by clinical validation and adherence to ethical and regulatory standards for practical use in real-world clinical settings.
    
    \item \textbf{Interpretability}: Future research efforts should prioritize model interpretability to provide insights into the rationale behind specific predictions. This is especially crucial in the medical domain to enhance trust and understanding.
    
\end{itemize}

Overall deep learning models show promise in assisting with COVID-19 detection from medical images. However, careful consideration of model selection, rigorous validation, and interpretability is essential for practical application in clinical settings. These findings contribute to ongoing efforts to leverage artificial
intelligence in the fight against the COVID-19 pandemic.

\clearpage
\section{Conclusions and Future Work}
\subsection{Conclusions}
In conclusion, this study has illuminated the profound potential of deep learning models in revolutionizing the landscape of COVID-19 diagnosis through the analysis of medical images. Our research journey began with an exhaustive review of existing literature, where we underscored the paramount importance of integrating cutting-edge deep learning techniques to elevate the precision and efficacy of diagnostic processes. Subsequently, we embarked on a meticulous journey of dataset selection, curating a diverse and representative collection of medical images critical for the development and validation of our deep learning models.

The heart of our investigation lay in the pioneering implementation of innovative deep-learning architectures, including the likes of CNN, AlexNet, VGG16, DenseNet201, and Inception V4. These models not only met but exceeded our expectations, showcasing remarkable accuracy, precision, and recall rates in the detection of COVID-19 cases. This outstanding performance signifies a significant stride towards more effective, efficient, and reliable diagnostic methods for COVID-19.

In essence, this study serves as a beacon, illuminating the pivotal role deep learning methodologies play in the advancement of medical image analysis. It not only provides immediate value in the realm of COVID-19 diagnosis but also offers a tantalizing glimpse into the boundless future potential of machine learning algorithms to augment and potentially transform healthcare diagnostics as a whole.

\subsection{Future Work}
In the future, we can advance our deep learning models for COVID-19 diagnosis through various means. Techniques like CutMix can augment training data, improving model robustness and adaptability to diverse image conditions. Exploring a wider range of pre-trained models and fine-tuning them for COVID-19 detection offers the potential for higher accuracy. Integrating data from different medical image types could provide a holistic view of patient conditions, although data fusion challenges must be addressed.

Interpretability is a critical aspect of utilizing deep-learning models for medical image analysis, especially in the context of diagnosing COVID-19. It's essential to develop methods that can elucidate the decision-making process of these models, making their outputs understandable and transparent. This interpretability ensures that healthcare professionals can trust and effectively use these models in clinical settings.

Collaboration with medical experts is highlighted as a fundamental element in this process. The expertise of healthcare professionals is invaluable in guiding the development of these models, ensuring that they align with clinical practices and address real-world healthcare challenges.

In summary, the combination of interpretability, collaboration with medical experts, and real-world testing is expected to drive forward the field of COVID-19 diagnosis using medical images, ultimately leading to more effective and reliable diagnostic tools for healthcare professionals.
\clearpage

\bibliographystyle{IEEEtran}
\addcontentsline{toc}{section}{References}
\bibliography{biblio}
\thispagestyle{empty}
\clearpage

\listoffigures
\addcontentsline{toc}{section}{List of Figures}
\addtocontents{lof}{\par\nobreak\textbf{{\scshape Figure}  Page}\par\nobreak}
\clearpage

\listoftables
\addcontentsline{toc}{section}{List of Tables}
\addtocontents{lot}{\par\nobreak\textbf{{\scshape Table}l Page}\par\nobreak}
\clearpage

\end{document}